\documentclass[letter,11pt]{article}
\usepackage{jheppub}
\usepackage{lineno}
\usepackage{braket}
\usepackage{latexsym}
\usepackage{enumitem}
\usepackage{relsize}
\usepackage{exscale}
\usepackage{amsmath,amssymb,amsthm,mathtools}
\usepackage{appendix}

\DeclareMathOperator{\Tr}{Tr}

\theoremstyle{definition}

\theoremstyle{plain}

\theoremstyle{remark}

\title{Spacetime topology and geometry in JT Gravity from non-crossing permutations}

\author[1]{Curtis T.~Asplund}
\author[1]{and Nicholas A.~Parrilla}
\affiliation[1]{Department of Physics \& Astronomy, San Jos\'e State University, One Washington Square, San Jos\'e, California, 95192-0106, USA}
\emailAdd{curtis.asplund@sjsu.edu}
\emailAdd{nicholas.parrilla@sjsu.edu}

\abstract{
We study how discrete parameters which enumerate non-crossing permutations become geometric quantities in JT gravity. Analyzing non-crossing permutations on the disk together with two classes of annular non-crossing permutations, we find that under suitable continuum scaling limits the number of marked boundary points becomes the thermal boundary length, while the number of through-connections becomes a geodesic length. In this limit, the enumerations reproduce the universal low-temperature behavior of the Schwarzian theory on the disk and the orientable and orientation-reversing double-trumpet geometries. We then relate these constructions to higher-boundary non-crossing diagrams, whose continuum enumeration reproduces the JT result obtained by attaching trumpets to the top-degree part of genus-zero Weil--Petersson volumes, suggesting that the same trumpet--enumeration parameter mapping extends beyond the planar case. Together, these results establish a direct connection between non-crossing permutation enumerations and Weil--Petersson volumes with trumpets attached, a construction that plays a central role in the gravitational path integral in JT gravity, and suggest that non-crossing diagrams provide a combinatorial route to computing multi-boundary partition functions.
}

\begin{document}
\maketitle
\flushbottom

\section{Introduction}

Diagrammatic expansions often appear in physics as computational bookkeeping devices, e.g., the cluster expansion in statistical mechanics. In theories of gravity, they can acquire a more literal geometric
meaning. For example, in 't Hooft's double-line notation, Feynman diagrams in a matrix model description become ribbon graphs, and the power of $N$ carried by a diagram is determined by the genus of the surface on which it can be drawn \cite{t_hooft_planar_1974-1,francesco_2d_1995}. This observation underlies the use of matrix models as sums over discretized two-dimensional surfaces and their applications to a variety of two-dimensional quantum gravity theories that began in the 1990s \cite{brezin_planar_1978-1,brezin_exactly_1990-1,douglas_strings_1990-1,
gross_nonperturbative_1990-1}. More generally, combinatorial objects have repeatedly been used to encode geometric information in quantum gravity, ranging from Young diagrams and replica permutations to chord diagrams and tensor networks \cite{corley_giant_2002,lin_bubbling_2004,gross_taylor_1993,
maxfield2023countingstatesmodelreplica,Bintanja:2026vkr,
marolf_transcending_2020,marolf_observations_2021, abdalla_gravitational_2025,berkooz_cordial_2024,budd2026doublescaledsykboundarymetrics,hayden_holographic_2016,akers_reflected_2021,akers_reflected_2022,akers_reflected_2024,bueller2024tensornetworksblackhole,cheng_random_2024,pastawski_holographic_2015,Swingle_2012,qi2018spacetimerandomtensornetworks}. These examples motivate the broader question of when the combinatorics arising in a quantum theory can be regarded as a discrete precursor of emergent geometry. In this paper, we study this question using a class of non-crossing diagrams that, as we show, arise naturally in holographic theories of two-dimensional gravity, specifically JT gravity \cite{jackiw_lower_1985,teitelboim_gravitation_1983, maldacena_large-n_1999}.

\begin{figure}[t!]
    \centering
    \includegraphics[width=0.72\linewidth]{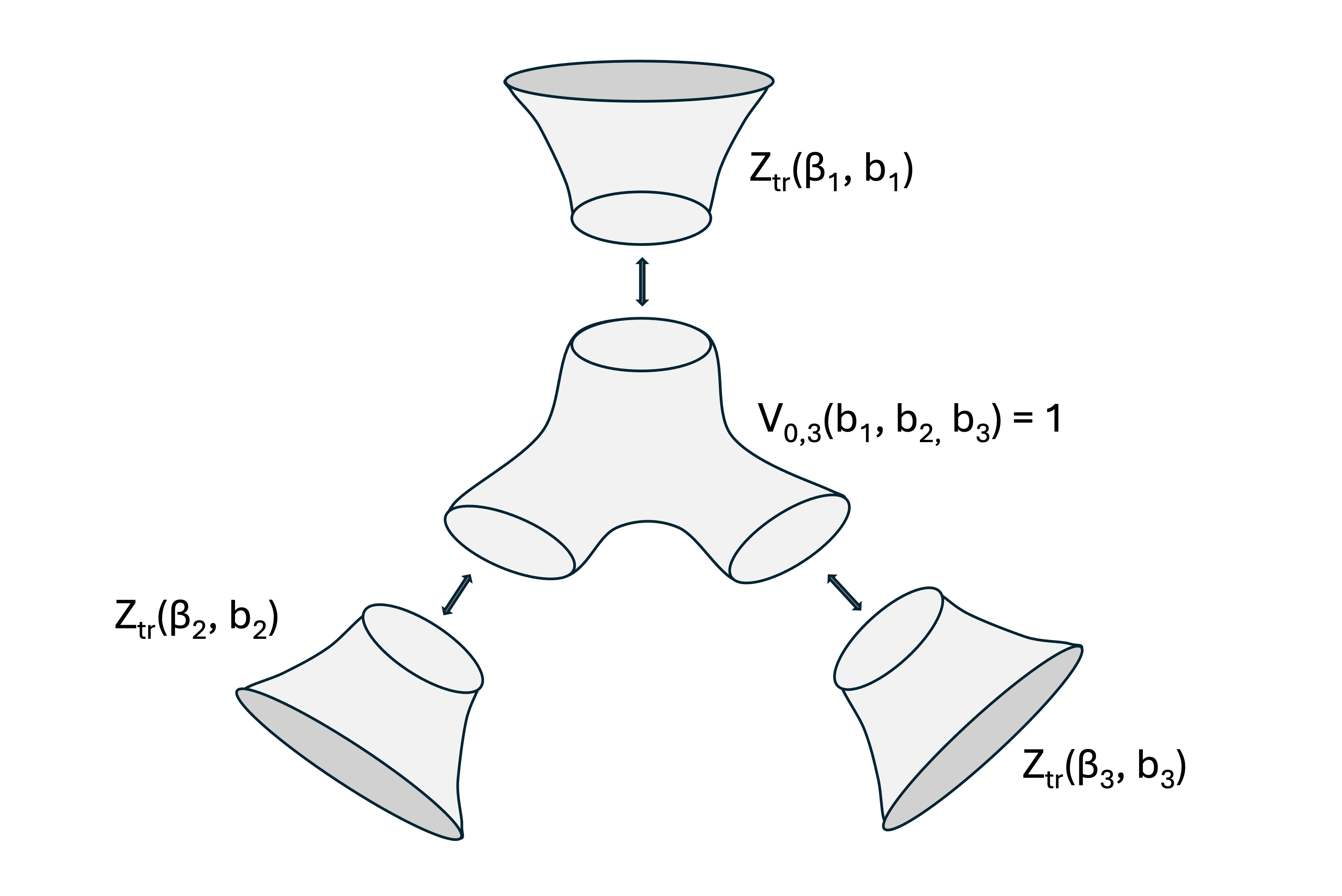}
    \caption{
    Decomposition of the genus-zero three-boundary JT geometry in the
    geodesic-length basis. Each asymptotic boundary of renormalized length
    $\beta_i$ is connected by a trumpet $Z_{\rm tr}(\beta_i,b_i)$ to a
    geodesic boundary of length $b_i$ on the pair-of-pants geometry. The
    three geodesic boundaries are integrated over according to
    Eq.~\eqref{eq:intro-gluing} and multiplied by the Weil--Petersson volume
    for the pair of pants, given by
    $V_{0,3}(b_1,b_2,b_3)=1$.}
    \label{fig:intro-gluing}
\end{figure}

JT gravity provides a particularly simple setting in which to study this question given its computational tractability as a theory of quantum gravity as well as it relation to random matrix theory
\cite{saad_jt_2019,stanford_jt_2020}. In Euclidean JT gravity, the dilaton constrains the curvature to $R=-2$, and the gravitational path integral reduces to a sum over two-dimensional hyperbolic surfaces \cite{saad_jt_2019, mertens_solvable_2023}. A surface of genus $g$ with $n$
boundaries carries the topological weight $e^{S_0\chi}$, with $\chi=2-2g-n$, and
so the expansion in $e^{-S_0}$ plays the role of the matrix-model expansion in $1/N$ \cite{saad_jt_2019, stanford_jt_2020}. At fixed topology, the remaining integral is over the moduli space of hyperbolic surfaces. For $n$ asymptotic boundaries of renormalized lengths $\beta_1,\ldots,\beta_n$, the multi-boundary partition function takes the form \cite{saad_jt_2019, mertens_solvable_2023}
\begin{equation}
    Z_{g,n}(\beta_1,\ldots,\beta_n)
    =
    e^{(2-2g-n)S_0}
    \int_0^\infty
    \prod_{i=1}^{n}b_idb_i
    V_{g,n}(b_1,\ldots,b_n)
    \prod_{i=1}^{n}
    Z_{\rm tr}(\beta_i,b_i)\ ,
    \label{eq:intro-gluing}
\end{equation}
where $V_{g,n}(b_1,\ldots,b_n)$ is the Weil--Petersson volume for the moduli space of genus-$g$ Riemann surfaces with $n$ geodesic boundaries of lengths $b_1,\ldots,b_n$, respectively, and
\begin{equation}
    Z_{\rm tr}(\beta,b)
    =
    \frac{1}{\sqrt{4\pi\beta}}
    e^{-b^2/(4\beta)}
    \label{eq:intro-trumpet}
\end{equation}
is the trumpet partition function joining an asymptotic boundary of length $\beta$ to a geodesic boundary of length $b$ \cite{saad_jt_2019, mertens_solvable_2023}.
Figure~\ref{fig:intro-gluing} illustrates this decomposition for the three-boundary case, where three trumpets are sewn to the geodesic boundaries of a pair-of-pants geometry.

We will focus first on the disk and the double trumpet. The disk is the leading genus-zero contribution to the one-boundary JT partition function. It consists of the topological factor $e^{S_0}$ multiplying the contribution of the Schwarzian boundary mode,
\begin{equation}
    Z_{\rm disk}(\beta)
    =
    e^{S_0}
    \frac{e^{2\pi^2/\beta}}{\sqrt{2\pi}\beta^{3/2}} \ , 
    \label{eq:intro-disk}
\end{equation}
which exhibits the universal $\beta^{-3/2}$ behavior at low
temperature. Similarly, the leading connected two-boundary geometry is the double trumpet,
\begin{equation}
Z_{\rm double\ trumpet}(\beta_1,\beta_2)
= \int_0^\infty b\ db\ Z_{\rm tr}(\beta_1,b) Z_{\rm tr}(\beta_2,b)\ ,
\label{eq:intro-double-trumpet}
\end{equation}
obtained by gluing two trumpets with boundary lengths $\beta_1$ and $\beta_2$ along a common geodesic of length $b$. The double trumpet is also closely related to the ramp in the connected spectral form factor, whose linear late-time growth is associated with random matrix spectral correlations and quantum chaos
\cite{mehta_random_1967,bohigas_characterization_1984,
cotler_black_2017,cotler_chaos_2017,hunter-jones_chaos_2018,
saad_semiclassical_2019,saad_jt_2019,saad_late_2019,
cipolloni_spectral_2023,saad_convergent_2024}. This relation to random matrix statistics reflects a more general and central role of random matrices in JT gravity, where the topological expansion of matrix integrals mirrors the genus expansion of the gravitational theory. One considers an integral over an ensemble of Hermitian matrices,
\begin{equation*}
\mathcal{Z}=\int dM\ e^{-N\Tr V(M)}\ .
\end{equation*}
Matrix integrals play a central role in JT gravity because their $1/N$ expansion is organized by genus, with the planar sector providing the leading contribution
\cite{t_hooft_planar_1974-1,francesco_2d_1995}. In the appropriate double-scaling limit, this genus expansion matches the topological expansion of JT gravity \cite{saad_jt_2019}. In this description, the thermal partition function is identified with the spectral observable
\begin{equation*}
Z(\beta)=\Tr e^{-\beta M}\ ,
\end{equation*}
and geometries with $n$ asymptotic boundaries are reproduced by connected correlators of $n$ such observables. In particular, the disk gives the leading one-point function $\langle Z(\beta)\rangle$, while the double trumpet gives the leading connected two-point function
$\langle Z(\beta_1)Z(\beta_2)\rangle_{\rm conn}$. More generally, these correlators satisfy
\begin{equation*}
\left\langle
\prod_{i=1}^{n}\Tr e^{-\beta_iM}
\right\rangle_{\text{conn}}
=
\sum_{g\geq0}
Z_{g,n}(\beta_1,\ldots,\beta_n)\ .
\end{equation*}
Thermal correlation functions can be expressed in terms of the
matrix-model eigenvalue density or, equivalently, in terms of matrix resolvents. The disk is determined by the leading spectral density, while higher-boundary and higher-genus contributions are encoded in its higher-point correlations. In the double-scaling limit, the matrix potential is tuned to the spectral edge appropriate to JT gravity, where the leading spectral density takes the form
\begin{equation}
    \rho_0(E) =
        \frac{\sinh\left(2\pi\sqrt{E}\right)}{4\pi^2}\ ,
\end{equation}
up to normalization. Near the lower spectral edge at $E=0$, $\rho_0(E)\sim\sqrt{E}$, and its Laplace transform gives the $\beta^{-3/2}$ behavior of the disk. Higher resolvents are generated by Eynard--Orantin topological recursion
\cite{eynard2007invariantsalgebraiccurvestopological}. For the JT spectral curve, this reproduces Mirzakhani's recursion for the Weil--Petersson volumes appearing in Eq.~\eqref{eq:intro-gluing}
\cite{saad_jt_2019}.

The geometric variables enter when we pass from the spectral
description to the geodesic-length basis in
Eq.~\eqref{eq:intro-gluing}, which is labeled by the geodesic lengths $b_i$. Given the history of diagrammatic expressions of similar matrix models in low-dimensional gravity, it is natural to ask whether JT gravity admits a comparable discrete description of these geometric variables. In this work, we argue that non-crossing permutations provide such a description. Non-crossing permutations appear frequently in free probability
\cite{voiculescu_limit_1991,2014arXiv1404.3393S,mingo_free_2017,
speicher_combinatorial_1998} where moments of random-matrix eigenvalue distributions are expressed as sums over non-crossing permutations. Connected fluctuations of two traces are similarly computed using annular non-crossing permutations, with marked points on each of the two boundary components and lines drawn connecting these points in a non-crossing fashion \cite{mingo_annular_2004,mingo_second_2007}.\footnote{
Closely related permutations also arise in replica calculations of entanglement entropy in replica-wormhole geometries
\cite{almheiri_replica_2020,penington_replica_2022,maxfield_bit_2022,
giddings_wormhole_2020,bousso_fluctuations_2023}.
}\footnote{ We discuss ideas connecting random matrix theory and diagrams appearing in free probability theory more thoroughly in Appendix~\ref{app:resolvent-rtransform}.
} Non-crossing diagrams also arise as the leading planar contributions to large-$N$ expansions of matrix integrals based on Wick contractions and the Weingarten calculus \cite{collins_weingarten_2022}. We show that their representations on disks, annuli, and more general surfaces suggest a connection with the two-dimensional geometries appearing in JT gravity. Under suitable scaling limits of the relevant discrete parameters, these enumerations reproduce continuum geometric quantities in JT gravity.

There are two closely related precedents for this viewpoint. Do, He, and Mathews studied non-crossing permutations on surfaces with arbitrary genus
and number of boundaries and showed that the leading terms in their enumeration are governed by intersection numbers, equivalently by the highest-degree part of Weil--Petersson volumes
\cite{do2019countingnoncrossingpermutationssurfaces}. Related connections between higher-order free probability, maps, and topological recursion have
been developed in \cite{borot_functional_2023}. Separately, on the physics side, Cao and Faulkner identified annular non-crossing diagrams arising in a replica calculation of the spectral form factor with a discretization of the double-trumpet geometry \cite{cao_ramp_2025}.\footnote{
Related permutation combinatorics appear in
\cite{kudler-flam_wormholes_2026,chen_chaos_2026}.
}
These results suggest complementary aspects of the same idea, where the enumeration of non-crossing diagrams have Weil--Petersson volumes as scaling limit,
while the diagrams themselves admit a direct interpretation in terms of JT geometries.

Our construction is also closely related to recent work on
double-scaled SYK (DS-SYK). Okuyama showed that connected DS-SYK correlators contain discrete analogues of Weil--Petersson volumes \cite{okuyama_discrete_2023}, while Do and Norbury proved that these recover ordinary Weil--Petersson volumes in the appropriate $q\to1$ limit \cite{do_norbury_dssyk_2025}. More generally, matrix correlators away from the double-scaling limit have recently been interpreted as discrete volumes obeying a discrete analogue of Mirzakhani recursion
\cite{giacchetto_discrete_2025}. These results show that discrete combinatorial quantities can have continuum limits governed by JT geometry. We ask a complementary question: can the discrete parameters appearing directly in non-crossing permutation enumerations be identified with the geometric variables entering JT partition functions? We make this discrete-to-continuum correspondence concrete by considering first the disk and then the annulus. For the disk, the
large-order enumeration reproduces the universal $\beta^{-3/2}$ behavior of the JT spectral edge. For the annulus, the numbers of boundary points scale to the asymptotic boundary lengths, while the number of through-connections scales to the internal geodesic length $b$. In this limit, the displaced binomial coefficients become Gaussian factors with the dependence characteristic of JT trumpet geometries.

This interpretation becomes natural after decomposing an annular non-crossing permutation into circular half-permutations \cite{kusalik_orthogonal_2005,cao_ramp_2025}.\footnote{A similar decomposition for chord diagrams in the DS-SYK model appears in \cite{Cao:2025pir}.} Each half-permutation contains a binomial factor that becomes a single trumpet partition function in the continuum limit, while the cyclic alignments of the two halves provide the discrete gluing multiplicity. Summing over the number of through-connections then produces the $b\,db$ measure in Eq.~\eqref{eq:intro-double-trumpet}. Allowing orientation-reversing identifications introduces a second relative-orientation set of diagrams, yielding the corresponding factor of two in the double-trumpet contribution of the real matrix model in the orthogonal universality class
\cite{stanford_jt_2020,mehta_random_1967,forrester_review_2022, weber_unorientable_2024}. Thus, the permutation description retains information absent from the underlying set partition, encoding not only continuum geometric variables but also information about the gluing operation and physical symmetries.

We extend the correspondence to connected genus-zero non-crossing permutations with an arbitrary number of boundaries. Their continuum limits reproduce the JT geometries obtained by attaching trumpet factors to the top-degree part of the genus-zero Weil--Petersson volumes. For three boundaries this gives the pair-of-pants geometry, with $V_{0,3}=1$, while the four-boundary case shows that the enumeration does not capture the constant term in $V_{0,4}$. This is
consistent with the result of
\cite{do2019countingnoncrossingpermutationssurfaces} that the leading polynomial coefficients of these diagram counts reproduce the top-degree Weil--Petersson data. Finally, a genus-one polygon enumeration provides evidence that the trumpet interpretation extends beyond planar surfaces. Together, these examples highlight which part of
the correspondence studied here is universal and where additional combinatorial information may be required to recover the lower-degree terms of the JT spectral curve.

The rest of the paper is organized as follows.
Section~\ref{section2} defines the combinatorial objects, considers the
one-boundary case, and shows that the large-$n$ limit of the non-crossing
count reproduces the universal disk scaling.
Section~\ref{section3} introduces annular non-crossing diagrams, their
decomposition into half-permutations, and the emergence of the trumpet and
double-trumpet gluing measure.
Section~\ref{nonorientable} extends the construction to annular
non-crossing permutations with reversed relative orientation.
Section~\ref{sec:higher-order-jt} studies the genus-zero higher-boundary enumeration, including the pair of pants and the emergence of the top-degree Weil--Petersson volumes, together with a first higher-genus example. 

\section{Non-crossing Partitions and the Disk}
\label{section2}

The simplest correspondence between non-crossing permutations and topologies and geometries appearing in JT gravity occurs between non-crossing permutations drawn on a disk and the JT hyperbolic disk geometry. We begin by studying how the universal low-temperature behavior of the JT disk is reflected in the combinatorics of these diagrams. The basic object consists of $n$ marked points arranged cyclically along a boundary. In an appropriate continuum limit, this discrete boundary will be identified with the asymptotic boundary of the JT disk, while the non-crossing conditions imposed on the diagrams provides combinatorial rules that will generalize to the annulus and higher-boundary surfaces.

Although our focus is combinatorial, non-crossing permutations also arise when analyzing the large-$N$ statistics of random matrices. We briefly develop this connection to make contact with the familiar random-matrix description of JT gravity. Throughout this section we use the complex Wishart ensemble, which was also the setting of Cao and Faulkner's study of the spectral form factor of random reduced density operators \cite{cao_ramp_2025}, where they also proposed that these non-crossing permutations resembled a discretization of the double trumpet geometry. The moments of the Wishart ensemble, whose eigenvalue distribution converges to the Marchenko--Pastur law in the large-$N$ limit \cite{marchenko_distribution_1967}, admit an expansion in noncrossing diagrams
\cite{mingo_free_2017,2014arXiv1404.3393S}. The particular ensemble, however, will not be essential to the continuum behavior we study. The square-root behavior of a regular soft spectral edge is shared by many random-matrix ensembles, including the Gaussian Unitary Ensemble (GUE), and also characterizes the low-energy edge of the JT spectral density. Consequently, the continuum limit considered here captures the universal soft-edge behavior of the JT disk, but not the additional information contained in the full JT spectral curve.

\subsection{Non-crossing Partitions and Permutations}
\label{sec:nc-defn}

We begin by distinguishing non-crossing permutations from non-crossing partitions. For points arranged on a single boundary, there is a bijection between the two, but this bijection does not in general extend to surfaces with multiple boundary components \cite{mingo_free_2017,mingo_annular_2004}. Consider $n$ points labeled $1,\ldots,n$ placed in cyclic order on a circle. A partition $\mathcal{P}$ of $\{1,\ldots,n\}$ into disjoint blocks is crossing if there exist $a<b<c<d$ such that $a$ and $c$ belong to one block while $b$ and $d$ belong to another, and non-crossing otherwise. We write ${\rm NC}(n)$ for the set of non-crossing partitions of $n$ points and $\#(\mathcal{P})$ for the number of blocks of $\mathcal{P}$. The non-crossing condition can be viewed geometrically, where a partition is non-crossing when its blocks can be represented inside the disk without intersections between distinct blocks. For example, every partition of three or fewer points is non-crossing and $|{\rm NC}(3)|=5$. The first exception occurs at $n=4$, where $\{\{1,3\},\{2,4\}\}$ cannot be drawn without a crossing, leaving $|{\rm NC}(4)|=14$ of the fifteen total set partitions that can be drawn. In general 
\begin{equation}
    |{\rm NC}(n)|=C_n = \frac{1}{n+1}\binom{2n}{n} \ ,
\end{equation}
the $n$th Catalan number, and a number that also appears when computing moments of random matrix eigenvalue distributions \cite{wigner_distribution_1958}.

While a partition records which points belong to the same block, a permutation additionally specifies a cyclic ordering of the elements within each block. On the disk, the additional information corresponds to the orientation of the boundary and is shown graphically by arrows representing the orientation of the permutations.\footnote{The cycle information is redundant on the disk since the boundary already encodes a label ordering.} Let $\gamma_n=(12\cdots n)$ be the cycle generating the cyclic order of the boundary points, and give each block the cyclic order it inherits from $\gamma_n$. That is, for every block of a non-crossing partition, its elements are ordered according to their order in $\gamma_n$. Then, the resulting block with ordering is regarded as a permutation cycle. The product of disjoint cycles acquired in this way form a unique permutation $\pi \in S_n$, where $S_n$ is the symmetric group of $n$ elements \cite{mingo_free_2017}. The non-crossing condition also extends to permutations. For a permutation $\sigma \in S_n$, define the permutation length as
\begin{equation}
    |\sigma| = n - \#(\sigma) \ ,
\end{equation}
where now $\#(\sigma)$ is the number of cycles in $\sigma$.\footnote{The permutation length $|\sigma|$ is the minimum number of transpositions whose composition gives $\sigma$. It defines the Cayley distance on $S_n$ by $d(\sigma,\tau)=|\sigma^{-1}\tau|$.} Since the boundary cycle $\gamma_n$ consists only of a single cycle, its permutation length is $|\gamma_n| = n - 1$. A permutation $\sigma$ is non-crossing when
\begin{equation}
    |\sigma| + |\sigma^{-1}\gamma_n| = |\gamma_n| \ ,
\end{equation}
called the geodesic condition \cite{mingo_free_2017}.\footnote{For an arbitrary intermediate permutation, the triangle inequality gives
$$ d({\rm id},\gamma_n)\leq d({\rm id},\sigma)+d(\sigma,\gamma_n). $$
The non-crossing condition saturates this inequality, meaning the route ${\rm id}\rightarrow\sigma\rightarrow\gamma_n$ is the shortest path from the identity to the boundary cycle on the Cayley graph of $S_n$, so $\sigma$ lies on a geodesic between them.} This non-crossing condition for permutations can also be written in terms of the number of its cycles, giving
\begin{equation}
    \#(\sigma) + \#(\sigma^{-1}\gamma_n) = n + 1 \ .
\end{equation}
Non-crossing permutations can be drawn as non-crossing partitions of the disk, with arrows indicating the cyclic orientation of each block. 

\begin{figure}
    \centering
    \includegraphics[width=0.95\linewidth,
    trim=0 0.5cm 0 0.5cm,
    clip]{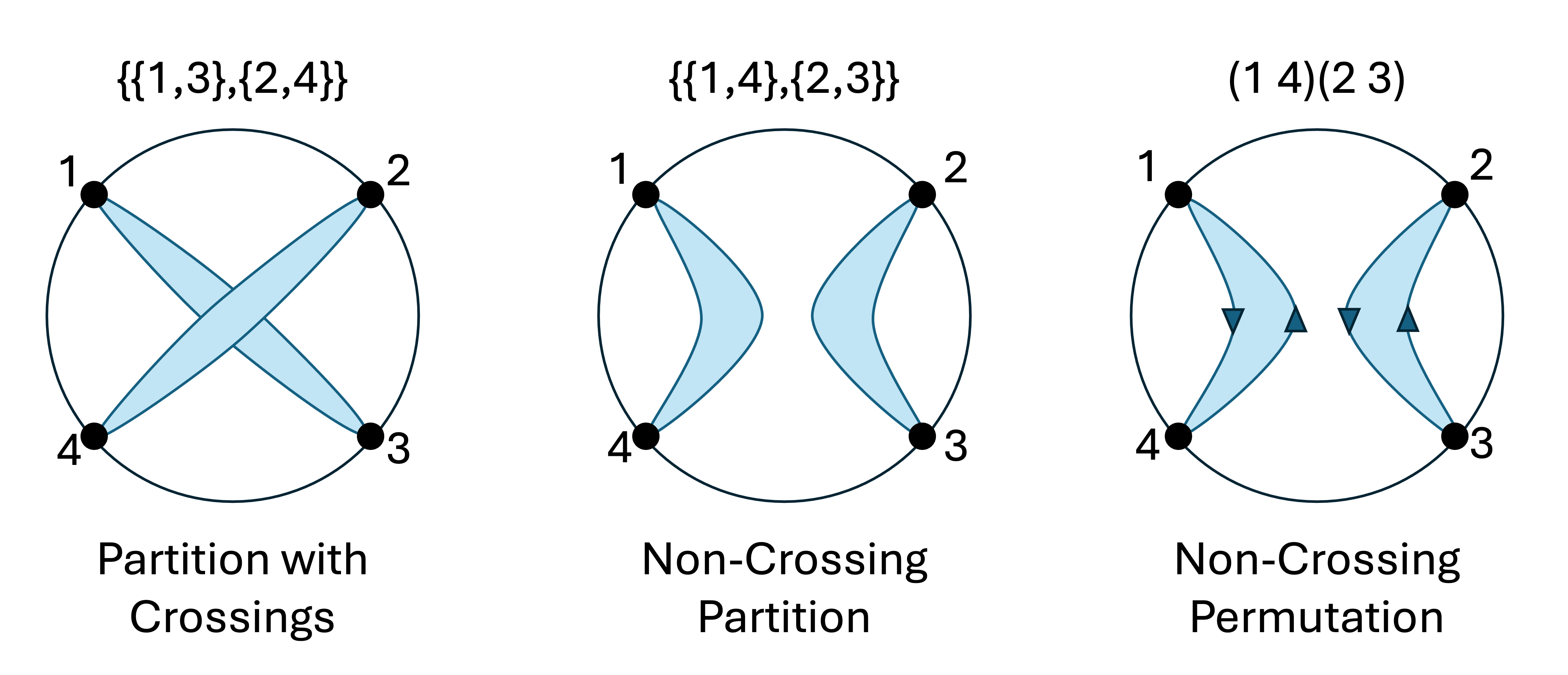}
    \caption{Examples of partitions and permutations on four marked boundary points. (Left) The partition $\{\{1,3\},\{2,4\}\}$ is crossing when drawn inside the disk. (Middle) $\{\{1,4\},\{2,3\}\}$ is a non-crossing partition.
    (Right) The same non-crossing partition is represented by the permutation $(14)(23)$, with arrows indicating the action of each
    cycle. We omit these arrows in later diagrams and use only the cycle notation.}
    \label{fig:NC4}
\end{figure}

Furthermore, the cycle-counting form of the non-crossing condition encodes the topology of the associated surface, with the relevant cycle counts determining its genus \cite{cori_hetvei_2013}.
\begin{equation}
\#(\sigma)+\#(\sigma^{-1}\gamma_n)=n+1-2g \ .
\end{equation}
Therefore, the geodesic condition is precisely the genus-zero case, $g=0$, corresponding to planar, non-crossing diagrams. The non-crossing condition can therefore be understood either pictorially, as the absence of crossings in a disk, or algebraically, through the permutation geodesic equation. This distinction becomes especially important when considering non-orientable surfaces, which we consider in Sec.~\ref{nonorientable}, where the relative orientations of permutation cycles and the possibility of orientation-reversing identifications provide additional data that cannot be captured by just the partitions. Figure~\ref{fig:NC4} illustrates the difference between partitions and permutations with and without crossings.

\subsection{Large-$N$ Matrix Moments from Non-Crossing Permutations}

We now turn from the individual non-crossing permutations to their enumeration, highlighting how diagram counts arise in computations of matrix moments in the large-$N$ limit, providing a direct connection between the diagrams introduced above and the random-matrix description of JT gravity. Here, we highlight the connection between non-crossing permutations and the complex Wishart ensemble, for which a member of the ensemble can be written as\footnote{We could have also studied the Gaussian Unitary Ensemble which is in the same universality class as the Wishart ensemble. In fact, both ensembles share the same scaling of the spectrum soft edge. We choose the Wishart ensemble since the order of moment studied corresponds to the number of marked points on the boundary of the non-crossing permutation, making the connection between moments and diagrams especially clear.}
\begin{equation}
    W=\frac{1}{N}X^\dagger X\ ,
    \label{eqn:Wishart}
\end{equation}
where $X$ is an $M\times N$ matrix with independent complex Gaussian entries of zero mean and unit variance. Square Wishart matrices arise, for example, as reduced density operators. 
In the limit
\begin{equation*}
    M,N\rightarrow\infty\ ,
    \qquad \text{and} \qquad
    \frac{M}{N}\rightarrow\lambda \ ,
\end{equation*}
the limiting moments of the eigenvalue distribution of $W$ can be written as

\begin{equation}
    m_n
    =
    \lim_{N\rightarrow\infty}
    \mathbb{E}\left[
        \frac{1}{N}\Tr(W^n)
    \right]
    \label{eqn:WishartMoment}
\end{equation}
and they admit an expansion in terms of non-crossing permutations, written formally as \cite{mingo_free_2017}
\begin{equation}
    m_n
    =
    \sum_{\pi\in{\rm NC}(n)}
    \lambda^{\#(\pi)} \ .
    \label{eqn:freepoisson}
\end{equation}
So the planar, non-crossing condition corresponds to the large-$N$ limit of the matrix model. Grouping the permutations according to their number $k$ of cycles gives
\begin{equation}
    m_n
    =
    \sum_{k=1}^{n}
    \frac{1}{n}
    \binom{n}{k}
    \binom{n}{k-1}
    \lambda^k\ ,
    \label{eqn:Nar}
\end{equation}
where
\begin{equation}
    N(n,k) =
    \frac{1}{n}
    \binom{n}{k}
    \binom{n}{k-1}
\end{equation}
are the Narayana numbers \cite{speicher_combinatorial_1998, mingo_free_2017}. 

For square Wishart matrices, $\lambda=1$, and so every non-crossing permutation receives equal weight and Eq.~\eqref{eqn:freepoisson} reduces to
\begin{equation}
    m_n = \sum_{k=1}^{n}
        \frac{1}{n}
        \binom{n}{k}
        \binom{n}{k-1} = \frac{1}{n+1}\binom{2n}{n} = C_n \ ,
\end{equation}
where $C_n$ is the $n$th Catalan number. We note that the Catalan numbers also appear in the GUE, where $C_{2n}$ counts the non-crossing pairings contributing to the $2n$-th moments and the odd moments vanish. We use square Wishart matrices because the $n$th moment is related directly to a non-crossing diagram with $n$ boundary points. We highlight that the different limits considered here have both diagrammatic and spectral significance in the random matrix model. The large-$N$ limit leads to the non-crossing permutation expansion of the matrix moments. On the other hand, the large-$n$ limit, which we consider next, probes the continuum behavior of the permutation enumeration and corresponding diagrams while probing the soft edge of the eigenvalue spectrum.

\subsection{Large-$n$ Limit of the Disk Enumeration}
\label{sec:universal-limit}

Here, we study the asymptotic scaling of the Catalan number counting the number of non-crossing permutations on the disk. The key result for this section is the large-$n$ behavior that follows directly from Stirling's approximation,
\begin{equation}
    C_n
    =
    \frac{(2n)!}{(n+1)(n!)^2}
    \sim
    \frac{4^n}{\sqrt{\pi}n^{3/2}} \ .
    \label{eqn:CatAsym}
\end{equation}
The factors appearing here can be related to the characteristics of the eigenvalue distribution of the corresponding matrix model. The factor $4^n$ depends on the normalization and location of the spectral edge, while the power $n^{-3/2}$ follows from the square-root behavior of the density near that edge. This can be seen explicitly when writing the matrix moments in terms of the spectral density. The limiting moments in Eq.~\eqref{eqn:WishartMoment} can equivalently be written as
\begin{equation}
    m_n
    =
    \int dx\rho(x)x^n \ .
    \label{eqn:spectralMoment}
\end{equation}
For the square Wishart ensemble, the large-$N$ eigenvalue density is the Marchenko--Pastur law,
\begin{equation}
    \rho_{\rm MP}(x)
    =
    \frac{1}{2\pi}\sqrt{\frac{4-x}{x}} \ ,
    \qquad
    x\in[0,4] \ .
    \label{eqn:MPDensity}
\end{equation}
So the connection between non-crossing diagrams and spectral laws is that at large-$N$ they compute the same moments,
\begin{equation}
    C_n
    =
    \sum_{\pi\in{\rm NC}(n)}1
    =
    \int_0^4 dx\rho_{\rm MP}(x)x^n \ .
    \label{eqn:diagramSpectralMoment}
\end{equation}
On non-crossing diagrams, $n$ is also the number of marked points on the boundary, so increasing $n$ reduces the separation of points and they approach a continuous circle and the non-crossing permutations fill in more of the disk. 

\begin{figure}
    \centering
    \includegraphics[width=0.95\linewidth]{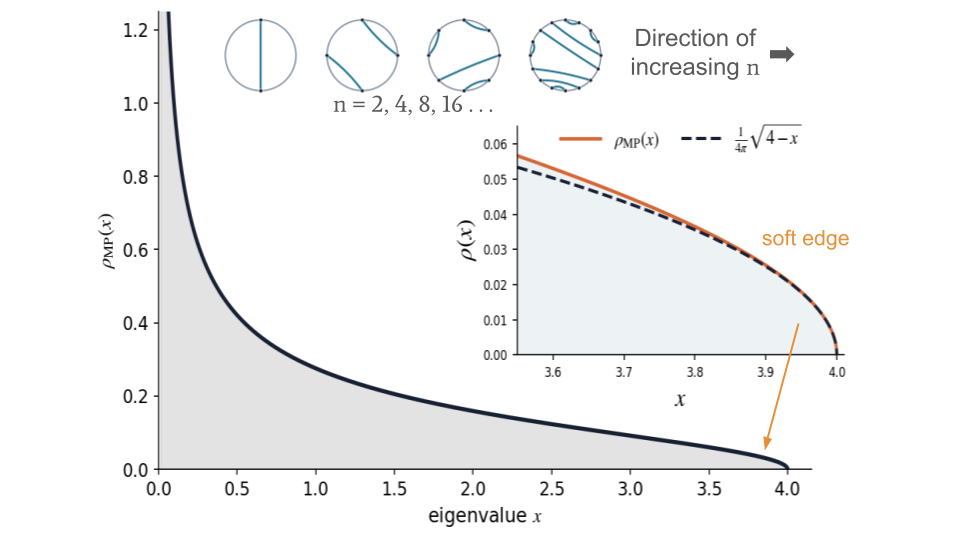}
    \caption{Higher moments probe the spectrum increasingly close to its upper edge.
    At the same time, the non-crossing diagrams contain more boundary points. The inset shows the Marchenko--Pastur density approaching its square-root behavior near the edge at $x=4$.
    }
    \label{fig:MPEdge}
\end{figure}

Likewise, higher-order moments in $n$ increasingly probe the spectrum near its upper edge. This connection between increasing $n$ in non-crossing partitions and the spectral edge is illustrated in Fig.~\ref{fig:MPEdge}. For the Marchenko--Pastur density, the upper edge lies at $x_\ast=4$, where
\begin{equation}
    \rho_{\rm MP}(4-\epsilon)
    \sim
    \frac{\sqrt{\epsilon}}{4\pi}
    \qquad \text{as} \qquad
    \epsilon\rightarrow0^+ \ .
    \label{eqn:MPEdge}
\end{equation}
The density therefore vanishes as a square root at the upper edge, which in general, occurs at regular soft edges in a broad class of random matrix ensembles.

To see how the edge behavior determines the large-$n$ scaling, suppose a limiting density is supported below an upper edge $x_\ast$ and behaves as
\begin{equation}
    \rho(x_\ast-\epsilon)
    \sim
    A\sqrt{\epsilon}
    \qquad \text{as} \qquad
    \epsilon\rightarrow0^+ \ .
    \label{eqn:generalSquareRootEdge}
\end{equation}
Near the edge, writing $x=x_\ast-\epsilon$, the factor appearing in the moment becomes
\begin{equation}
    x^n
    =
    x_\ast^n
    \left(1-\frac{\epsilon}{x_\ast}\right)^n
    \sim
    x_\ast^n
    \exp\left(-\frac{n\epsilon}{x_\ast}\right) \ .
    \label{eqn:largeMomentExponential}
\end{equation}
The large moments are therefore
\begin{align}
    m_n
    &\sim
    x_\ast^n
    \int_0^\infty d\epsilon\ A\sqrt{\epsilon}
    \exp\left(-\frac{n\epsilon}{x_\ast}\right)
    \nonumber\\
    &=
    A\ \Gamma\left(\frac{3}{2}\right)
    x_\ast^{n+3/2}n^{-3/2} \ .
    \label{eqn:generalMomentEdge}
\end{align}
The factor $x_\ast^n$ reflects the position of the spectral edge, whereas the power $n^{-3/2}$ depends only on its square-root behavior and is universal. For the Marchenko--Pastur density, $x_\ast=4$ and $A=1/4\pi$, giving
\begin{equation}
    m_n
    \sim
    \frac{4^n}{\sqrt{\pi}n^{3/2}},
\end{equation}
in agreement with Eq.~\eqref{eqn:CatAsym}.

To isolate the part of the diagram count that does not depend on the location of the spectral edge, we remove the non-universal factor $4^n$ and define
\begin{equation}
    \widetilde{C}_n
    =
    \frac{C_n}{4^n}
    \sim
    \frac{1}{\sqrt{\pi}n^{3/2}}.
    \label{eqn:normalized_cat}
\end{equation}
Equivalently, the spectrum can also be rescaled such that the relevant edge lies at $x_\ast=1$. Such a rescaling changes the exponential growth of the moments but not the universal scaling of the edge. Hence, the continuum behavior of the non-crossing enumeration is controlled by the same square-root edge behavior that appears in the JT spectral density.

\subsection{Comparison with the Disk Topology}

\begin{figure}[t]
    \centering
    \includegraphics[width=1\linewidth, trim=0 3.5cm 0 1.45cm, clip]{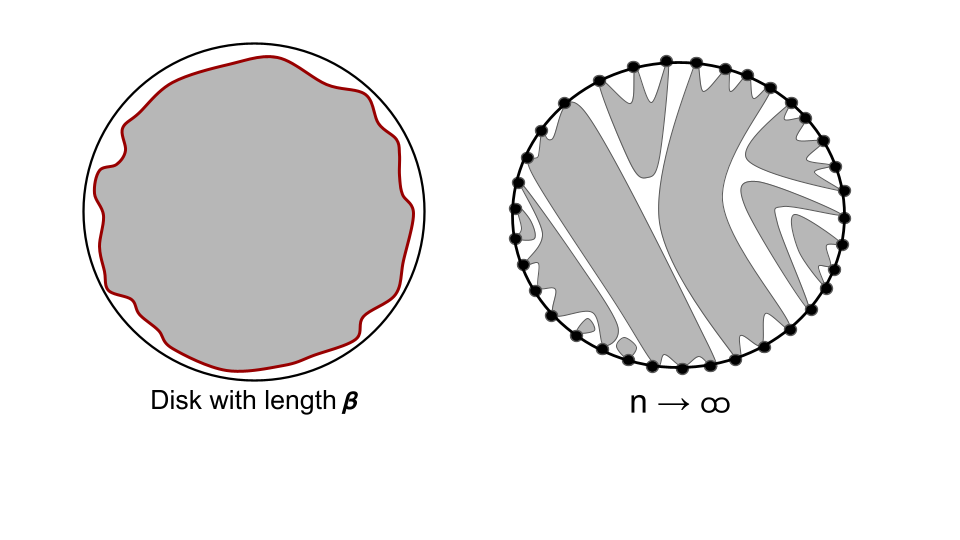}
    \caption{
    The disk geometry and its non-crossing discretization. The JT boundary of
    length $\beta$ is shown in red. The $n$ marked points of the non-crossing
    diagram provide a discrete boundary, with increasing $n$ probing the continuum
    limit. Under the scaling $n\leftrightarrow\beta$, the discrete boundary is
    identified with the thermal boundary of the JT disk.
    }
    \label{fig:DiskDict}
\end{figure}

We now compare the continuum behavior of the non-crossing enumeration with term in the partition function expansion associated with the disk topology in JT Gravity. Including its topological weight, the disk partition function is \cite{saad_jt_2019}
\begin{equation}
    Z_{\rm disk}(\beta)
    =
    e^{S_0}
    \frac{e^{2\pi^2/\beta}}
    {\sqrt{2\pi}\beta^{3/2}} \ ,
    \label{eqn:JTDisk}
\end{equation}
where $\beta$ is the renormalized length of the asymptotic boundary. At low temperature,
\begin{equation}
    e^{-S_0}Z_{\rm disk}(\beta)
    \sim
    \frac{1}{\sqrt{2\pi}\beta^{3/2}},
    \qquad
    \beta\rightarrow\infty.
    \label{eqn:JTDiskLowTemp}
\end{equation}
Eq.~\eqref{eqn:JTDiskLowTemp} and Eq.~\eqref{eqn:normalized_cat} appear formally similar, and we propose the identification
\begin{equation}
    n\longleftrightarrow\beta \ ,
    \label{eqn:DiskDictionary}
\end{equation}
up to an overall normalization of the variables, giving the first example of a correspondence between enumerative parameters of the non-crossing permutations and geometric parameters in JT gravity. The identification in Eq.~\eqref{eqn:DiskDictionary} can be understood as a continuum scaling relation where the continuous geometries emerge from a continuum limit of non-crossing permutations. Let $a$ be the ``lattice spacing" between marked boundary points and consider the limit $n\rightarrow\infty$ and $a\rightarrow0$ while the length is fixed at the value $\beta = an$. We can also visualize the cycles connecting points as filling the bulk geometry, as we illustrate in Fig. \ref{fig:DiskDict}.

The relation between $n$ and $\beta$ is also apparent in their respective spectral representations. Equation~\eqref{eqn:largeMomentExponential} shows that near the matrix edge the large power $x^n$ becomes
$    x^n
    \sim
    x_\ast^n
    e^{-n\epsilon/x_\ast},
$
while the thermal partition function weights the JT spectrum by
$
    e^{-\beta E}.
$
Thus, near the spectral edge, the moment order $n$ enters the matrix integral in the same way that the inverse temperature $\beta$ enters the thermal partition function. The Wishart moment probes the upper edge at $x=4$, while the JT integral probes the lower edge at $E=0$, but in both cases the density vanishes as a square root. Integrating this square-root behavior against the corresponding exponential weight gives the same universal scaling.

The disk provides the simplest example of the correspondence because it contains only a single boundary parameter. The annulus provides a stronger test because its non-crossing permutations contain additional enumerative information describing how the two boundaries are connected. In the next section, we study diagrams that also appear in computations of connected two-point functions involving powers of traces of random matrices. We will show that, in an analogous continuum scaling limit, the boundary point numbers become $\beta_1$ and $\beta_2$ of the double trumpet geometry, while the number of through-connections becomes the geodesic length $b$ along which the two trumpets are glued.

\section{Annular Non-crossing Diagrams and the Double Trumpet}
\label{section3}

Annular non-crossing (ANC) permutations extend the idea of non-crossing permutations to diagrams with two boundaries, which have $n$ and $m$ points on their two boundaries with $c$ through cycles connecting them. Annular non-crossing diagrams were previously shown to compute the ramp of the spectral form factor in a random matrix model \cite{cao_ramp_2025,kudler-flam_wormholes_2026}, which in JT gravity is computed via the double-trumpet geometry  \cite{saad_comments_2021, saad_convergent_2024}.

\subsection{Annular Non-crossing Permutations}
\label{sec:anc-defn}

The two-boundary analogue of Section~\ref{sec:nc-defn} replaces the disk by an annulus. Place $n$ points in cyclic order on one boundary circle and $m$ points in cyclic order on the other, and let 
\begin{equation}
    \gamma_{n,m}
    =
    (12\cdots n)(n+1\cdots n+m)
    \label{eqn:annular-boundary-permutation}
\end{equation}
be the permutation whose two cycles record these cyclic orders. A permutation $\tau$ of the $n+m$ points is an \emph{annular non-crossing permutation} when its cycles can be drawn without crossings in the annulus and when it connects the two boundaries, meaning that at least one cycle contains points from both circles. Equivalently, in terms of the cycle partitions associated with the
permutations, the
annular non-crossing condition is \cite{mingo_annular_2004,mingo_second_2007}
\begin{equation}
    \#(\tau)+\#(\tau^{-1}\gamma_{n,m})=n+m .
    \label{eqn:annular-conditions}
\end{equation}
We denote the set of permutations satisfying this condition by
${\rm ANC}(n,m)$.

For ANC permutations, the distinction between non-crossing partitions and permutations becomes important as there is not a bijection between the two. On the annulus, a cycle connecting both circles can have inequivalent cyclic orderings corresponding to different relative arrangements of the two boundaries.\footnote{ For example, consider $n=2$ and $m=1$. Three partitions of $\{1,2,1'\}$ connect the two boundaries,
\begin{equation}
    \{\{1,1'\},\{2\}\},
    \qquad
    \{\{2,1'\},\{1\}\},
    \qquad
    \{\{1,2,1'\}\}.
\end{equation}
However, the final partition corresponds to two annular non-crossing permutations,
\begin{equation}
    (121'),
    \qquad
    (11'2),
\end{equation}
which satisfy Eq.~\eqref{eqn:annular-conditions} separately and cannot be deformed into one another while preserving the annular embedding. There are therefore four annular non-crossing permutations rather than three connected non-crossing partitions.}

The annulus also introduces a new enumerative parameter. We call a cycle containing points on both boundary circles a \emph{through-cycle}, and denote the number of such cycles by $c$. An ANC permutation is connected if there is at least one through-cycle connecting points on the outer boundary to points on the inner boundary, that is, $1\leq c\leq\min(n,m)$. Here, $c$ is less than or equal to the number of boundary points since there can be at most as many through connections as there are boundary points. Once $c$ is fixed, the remaining cycles lie entirely on one of the two boundaries and are non-crossing there. Each side of the annulus therefore resembles a non-crossing partition of a disk, while the two sides are coupled through the number of through-cycles.

\begin{figure}
    \centering
    \includegraphics[width=1\linewidth, trim=0 3.5cm 0 4.5cm, clip]{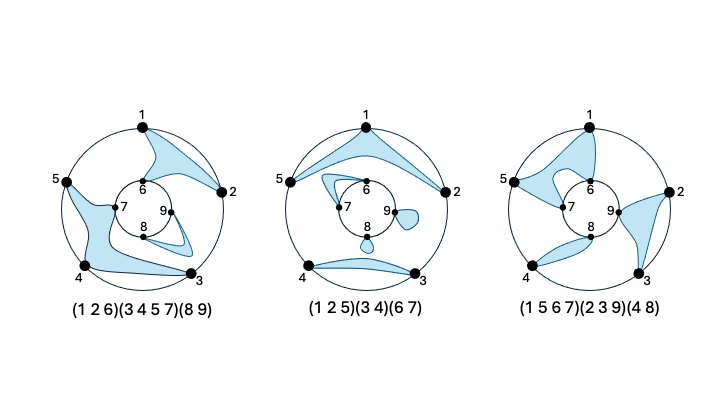}
    \caption{
    Examples of annular non-crossing permutations on five outer and four inner boundary points. (Left) The permutation $(1,2,6)(3,4,5,7)(8,9)$ contains two through-cycles, $(1,2,6)$ and $(3,4,5,7)$, connecting the outer and inner boundaries, together with the inner cycle $(8,9)$. (Center) The permutation $(1,2,5)(3,4)(6,7)$ contains two outer cycles, $(1,2,5)$ and $(3,4)$, and an inner cycle $(6,7)$, with $8$ and $9$ fixed. (Right) The permutation $(1567)(239)(48)$ consists entirely of
    through-cycles, each containing points from both boundaries. The shaded regions indicate the cycles of each permutation.
    }
    \label{fig:ANC_examples}
\end{figure}

This is a discrete version of an analogous rule in the double trumpet. At fixed geodesic length $b$, the two trumpet amplitudes depend separately on $\beta_1$ and $\beta_2$ and are coupled only through the common gluing geodesic. The remaining relative twist produces the measure associated with gluing along that geodesic. We will see that the refined annular enumeration has precisely the same factorized structure in the continuum limit. An example set of ANC permutations is shown in Fig. \ref{fig:ANC_examples}.

\subsection{Connected Two-Trace Moments}

The ANC permutations, like their disk counterparts, also play a role in analyzing random matrix statistics in the large-$N$ limit \cite{mingo_annular_2004, mingo_second_2005}. The matrix observable associated with the annulus is the connected correlator of two traces,
\begin{equation}
    \left\langle \Tr(W^n)\Tr(W^m) \right\rangle_{\text{conn}}
    =
    \left\langle \Tr(W^n)\Tr(W^m)\right\rangle
    -
    \left\langle \Tr(W^n)\right\rangle
    \left\langle\Tr(W^m)\right\rangle \ ,
    \label{eqn:connected-def}
\end{equation}
where $W$ is a complex Wishart matrix. This measures the covariance of two spectral observables, and the connected two-point function describes fluctuations about the limiting eigenvalue density\footnote{At a regular soft edge of a unitary Dyson universality class ($\beta_D=2$) random-matrix ensemble \cite{dyson_statistical_1962},
the edge fluctuations are universal, which means that after rescaling, the largest eigenvalue is governed by the Tracy--Widom distribution
TW$_2$. The local edge correlations are described by the Airy kernel
\cite{Tracy:1992kc,tracy_level-spacing_1994}.
}. The connected correlator can be written as a sum over annular non-crossing permutations \cite{mingo_annular_2004,mingo_second_2007,george_second_2025},
\begin{equation}
    \left\langle \Tr(W^n)\Tr(W^m)\right\rangle_{\text{conn}}
    =
    \sum_{\tau\in{\rm ANC}(n,m)}
    \lambda^{\#(\tau)} \ ,
    \label{eqn:anc-sum}
\end{equation}
where $\#(\tau)$ is the number of cycles of $\tau$ and $\lambda=\lim_{N,M \to \infty} M/N$.\footnote{The Wishart matrix has the same sum over permutations appearing for both its moments and connected correlation functions. In free probability theory, this is referred to the Free Poisson law.} 

\subsection{Enumeration by Through-Cycles}
\label{sec:annular-count}

The enumeration of annular non-crossing permutations can be organized according to the number $c$ of through-cycles. In addition to these through-cycles, an annular non-crossing permutation may contain $r$ cycles confined to the outer boundary and $s$ cycles confined to the inner boundary. The enumeration for computing the connected two point function can be written as a sum over permutations with $c$ through cycles, given by
\begin{equation}
    \left\langle \Tr(W^n)\Tr(W^m)\right\rangle_{\mathrm{conn}}
    =
    \sum_{c=1}^{\min(n,m)}
    \mathcal{N}(n,m,c;\lambda),
    \label{eqn:group-by-c}
\end{equation}
where
\begin{align}
    \mathcal{N}(n,m,c;\lambda)
    &=
    \sum_{r=0}^{n-c}
    \sum_{s=0}^{m-c}
    \#\mathrm{ANC}(n,m;c,r,s)\,
    \lambda^{c+r+s} \notag\\
    &=
    c\lambda^c
    \sum_{r=0}^{n-c}
    \sum_{s=0}^{m-c}
    \binom{n}{r}
    \binom{m}{s}
    \binom{n}{r+c}
    \binom{m}{s+c}
    \lambda^{r+s}.
    \label{eqn:anc-count-binomial}
\end{align}
Here $\#\mathrm{ANC}(n,m;c,r,s)$ denotes the number of annular
non-crossing permutations with $c$ through-cycles, $r$ outer-only cycles, and $s$ inner-only cycles
\cite{mingo_annular_2004,kim2012cyclicsievingphenomenonannular}. Equivalently, the sums over $r$ and $s$ factorize and can be repackaged via \cite{cao_ramp_2025}
\begin{equation}
    \mathcal{N}(n,m,c)
    =
    c\lambda^{c}
    \binom{n}{c}\binom{m}{c}
    {}_2F_1\!\left(-n,-n+c;c+1;\lambda\right)
    {}_2F_1\!\left(-m,-m+c;c+1;\lambda\right) \ ,
    \label{eqn:anc-count}
\end{equation}
where ${}_2F_1$ is the Gauss hypergeometric function. The formal structure of this expression already resembles the double trumpet, with two boundaries and a through-connection joining them. At fixed $c$, one factor depends only on $n$ and the other only on $m$, while the two are coupled through their common value of $c$. The prefactor $c$ counts the relative cyclic alignments of the through-cycles on the two boundaries.

As in the disk calculation, we specialize to square Wishart matrices, setting $\lambda=1$, so that the cycle weights are trivial. Since the first argument of each hypergeometric function is a negative integer, the series
\begin{equation}
    {}_2F_1(a,b;d;1)
    =
    \sum_{k\geq0}
    \frac{(a)_k(b)_k}{(d)_kk!}
\end{equation}
terminates. In the first boundary factor, $(-n+c)_k$ vanishes first, so the series terminates at $k=n-c$. The Chu--Vandermonde identity,
\begin{equation}
    {}_2F_1(-n,b;d;1)
    =
    \frac{(d-b)_n}{(d)_n} \ ,
\end{equation}
applies with $b=c-n$ and $d=c+1$, giving
\begin{equation}
    {}_2F_1(-n,-n+c;c+1;1)
    =
    \frac{(n+1)_n}{(c+1)_n} \ .
\end{equation}
Using
\begin{equation}
    (n+1)_n=\frac{(2n)!}{n!} 
    \qquad \text{and} \qquad
    (c+1)_n=\frac{(n+c)!}{c!} \ ,
\end{equation}
the complete boundary factor becomes
\begin{equation}
    \binom{n}{c}
    {}_2F_1(-n,-n+c;c+1;1)
    =
    \frac{(2n)!}{(n+c)!(n-c)!}
    =
    \binom{2n}{n-c}\ .
    \label{eqn:binomial-collapse}
\end{equation}
The same holds for the second boundary, so the ANC permutation enumeration reduces to 
\begin{equation}
    \mathcal{N}(n,m,c)
    =
    c
    \binom{2n}{n-c}
    \binom{2m}{m-c} \ ,
    \label{eqn:anc-simplified}
\end{equation}
which matches the known enumeration for ANC($n,m,c$) \cite{kim2012cyclicsievingphenomenonannular}.

\subsection{Continuum Limit and the Double Trumpet}

We now take the continuum limit of the enumeration in Eq.~\eqref{eqn:anc-simplified}. Here, the annulus contains the additional parameter $c$, which must be scaled together with the boundary sizes $n$ and $m$ in order to survive in the limit.

The relevant scaling of $c$ follows directly from the binomial factors where, for large $n$, the displaced central binomial coefficient $\binom{2n}{n-c}$ decays on the scale $c\sim\sqrt{n}$, and similarly for $m$. Including the prefactor $c$, Eq.~\eqref{eqn:anc-simplified} is maximized at
\begin{equation}
    c_\ast^2
    =
    \frac{nm}{2(n+m)},
\end{equation}
which gives $c_\ast=\sqrt{n}/2$ for equal boundary sizes $n=m$. We therefore take $n,m,c\rightarrow\infty$ with $\frac{c}{\sqrt{n}}$ and $\frac{c}{\sqrt{m}}$ held fixed. In this regime, the de Moivre--Laplace approximation leads to the asymptotic Gaussian behavior. To see this, write
\begin{equation}
    \binom{2n}{n-c} =
    \frac{(2n)!}{(n-c)!(n+c)!}.
\end{equation}
Using Stirling's approximation with $c=O(\sqrt n)$ gives \cite{papoulis2002probability, spencer2014asymptopia}
\begin{align}
    \log \binom{2n}{n-c}
    &=
    n\log 4
    -\frac{1}{2}\log(\pi n)
    -\frac{c^2}{n}
    +O\!\left(
        \frac{c^4}{n^3}+\frac{1}{n}
    \right) \ .
\end{align}
Since $c=O(\sqrt n)$ implies $c^4=O(n^2)$, the term
$c^4/n^3$ is $O(1/n)$, as are the remaining subleading corrections in Stirling's approximation, so the approximation becomes exact as $n\to\infty$. Exponentiation yields, in the continuum limit,
\begin{equation}
    \binom{2n}{n-c}
    \sim
    \frac{4^n}{\sqrt{\pi n}}
    e^{-c^2/n} \ ,
    \label{eqn:dml}
\end{equation}
and similarly for the second boundary. Substituting into Eq.~\eqref{eqn:anc-simplified} gives
\begin{equation}
    \mathcal{N}(n,m,c)
    \sim
    \frac{4^{n+m}}{\pi\sqrt{nm}}
    c
    e^{-c^2/n}
    e^{-c^2/m}\ .
    \label{eqn:anc-continuum}
\end{equation}
The two boundary factors have therefore become independent Gaussians coupled only through their common value of $c$. This is the discrete counterpart of cutting the double trumpet along its minimum geodesic where, at fixed gluing length, the two trumpets factorize, and all dependence between the two sides is carried by that common length. We highlight this further in the next subsection. In the continuum limit, the discrete sum may be replaced by an integral,
\begin{equation}
    \left\langle
        \Tr(W^n)\Tr(W^m)
    \right\rangle_{\text{conn}}
    \sim
    \frac{4^{n+m}}{\pi\sqrt{nm}}
    \int_0^\infty
    c\ dc \ 
    e^{-c^2/n}
    e^{-c^2/m}\ .
    \label{eqn:anc-integral}
\end{equation}
As in the disk calculation, the factor $4^{n+m}$ records the location of the Wishart spectral edge and can be removed by rescaling the spectrum. The remaining expression contains both the Gaussian dependence and the integration measure needed for the double trumpet. Using the same boundary-length convention as in Eq.~\eqref{eqn:DiskDictionary}, we identify
\begin{equation}
    n\longleftrightarrow\beta_1\ ,
    \qquad
    m\longleftrightarrow\beta_2\ ,
    \qquad
    c\longleftrightarrow\frac{b}{2}\ .
    \label{eqn:DTDictionary}
\end{equation}
The first two identifications are the direct two-boundary extension of the disk dictionary. The third follows by matching the terms in the Gaussians. Moreover, each displaced binomial coefficient becomes the Gaussians appearing in the JT trumpet in this continuum limit.

The combinatorial prefactor also has geometric interpretation. The factor $c$ in Eq.~\eqref{eqn:anc-simplified} counts the cyclic relative alignments of the through-cycles on the two boundaries. By Eq.~\eqref{eqn:DTDictionary},
\begin{equation}
    \frac{c\ dc}{\pi\sqrt{nm}}
    =
    \frac{b\ db}{4\pi\sqrt{\beta_1\beta_2}}
    =
    b\ db
    \frac{1}{\sqrt{4\pi\beta_1}}
    \frac{1}{\sqrt{4\pi\beta_2}}\ .
    \label{eqn:DTMeasureMatch}
\end{equation}
Hence, the discrete multiplicity and sum over through-cycles therefore become the Fenchel--Nielsen gluing measure $b\ db$ together with the normalization factors of the two trumpets \cite{saad_jt_2019, stanford_jt_2020}. After removing of the nonuniversal edge factor and mapping parameters, Eq.~\eqref{eqn:anc-integral} can be written and integrated, giving
\begin{equation}
    \int_0^\infty
    bdb
    Z_{\rm tr}(\beta_1,b)
    Z_{\rm tr}(\beta_2,b)
    =
    \frac{1}{2\pi}
    \frac{\sqrt{\beta_1\beta_2}}
    {\beta_1+\beta_2}\ ,
    \label{eqn:DTexact}
\end{equation}
which is precisely the double-trumpet amplitude. 

Additionally, the same result can also be obtained by first performing the discrete sum over the number of through-cycles and then taking the continuum limit. Summing Eq.~\eqref{eqn:anc-simplified} gives the total number of connected annular non-crossing permutations \cite{kim2012cyclicsievingphenomenonannular},
\begin{equation}
    |{\rm ANC}(n,m)|
    =
    \sum_{c=1}^{\min(n,m)}
    c\binom{2n}{n-c}\binom{2m}{m-c}
    =
    \frac{nm}{2(n+m)}
    \binom{2n}{n}\binom{2m}{m} \ .
    \label{eqn:anc-total}
\end{equation}
Using the central-binomial asymptotic
\begin{equation}
    \binom{2n}{n}
    \sim
    \frac{4^n}{\sqrt{\pi n}}\ ,
\end{equation}
and similarly for $m$, gives
\begin{equation}
    |{\rm ANC}(n,m)|
    \sim
    \frac{4^{n+m}}{2\pi}
    \frac{\sqrt{nm}}{n+m}\ .
    \label{eqn:anc-total-asymptotic}
\end{equation}
After removing the same nonuniversal edge factor $4^{n+m}$ and using $n\longleftrightarrow\beta_1$ and
$m\longleftrightarrow\beta_2$, this becomes
\begin{equation}
    \frac{1}{2\pi}
    \frac{\sqrt{\beta_1\beta_2}}
    {\beta_1+\beta_2}\ ,
\end{equation}
again reproducing the double-trumpet amplitude in Eq.~\eqref{eqn:DTexact}. The two derivations contain different amounts of geometric information. Taking the continuum limit of the total annular count reproduces the integrated double-trumpet amplitude directly, while resolving the enumeration by $c$ before taking the limit additionally identifies the discrete precursor of the gluing geodesic and its measure. In this sense, the two enumerations allow us to either study the double trumpet integral, or by summing over the through cycles, compute it directly. Thus, the annulus provides a stronger version of the disk correspondence, where there is not only the boundary dependence, but also the variable over which the two geometries are glued, all of which emerge from non-crossing combinatorics.

Finally, the order in the topological expansions for the two descriptions also agrees. In the random matrix setting, the annular diagrams compute connected fluctuations of the entanglement spectrum. For normalized traces, these fluctuations first appear at order $1/N^2$, while the connected correlator of two unnormalized traces is of order $N^0$. This agrees with the matrix-model scaling
$N^{2-2g-q} = N^0$ for a genus-zero surface with $q=2$ boundaries\footnote{We use $q$ to denote the number of boundaries here to differentiate it from the number of outer boundary points, $n$.}. On the gravity side, the same topology has Euler characteristic
\begin{equation}
    \chi=2-2g-q=0
\end{equation}
and therefore carries the weight $  e^{\chi S_0}=1$ for two boundaries and $g=0$. The $1/N$ expansion and the JT topological expansion therefore assign the same order to the annular non-crossing diagrams and the double-trumpet geometry. This feature further highlights how the permutation definitions are intrinsically topological, encoding not only genus through the planarity condition, but also encoding the entire Euler characteristic appearing in the topological expansion of the GPI.

\begin{figure}[t!]
    \centering
    \includegraphics[width=1\linewidth, trim=0 5.5cm 0 3.5cm, clip]{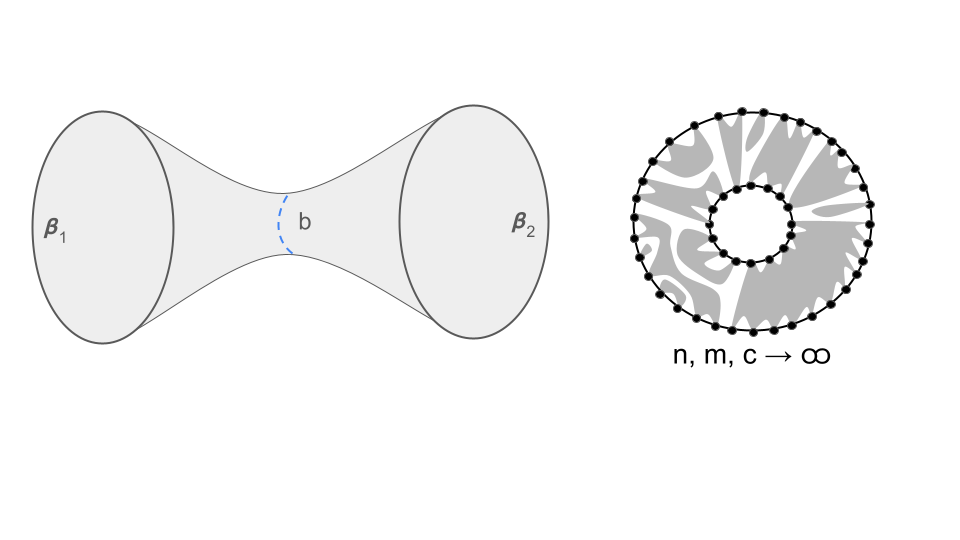}
    \caption{
    The JT double trumpet and its annular counterpart. On the left, two asymptotic boundaries of lengths $\beta_1$ and $\beta_2$ are joined along a geodesic of length $b$. On the right, an annular non-crossing permutation has $n$ and $m$ points on its boundaries and $c$ through-cycles connecting them. In the continuum limit, $n$ and $m$ become the thermal boundary lengths, while $2c$ becomes the gluing geodesic length. The multiplicity at fixed $c$ becomes the measure $b\ db$.
    }
    \label{fig:DiagramDuality}
\end{figure}

\subsection{Half-permutations and the Trumpet}

Here, we analyze the factorization of Eq.~\eqref{eqn:anc-simplified} into its individual trumpets and the gluing that creates the double trumpet from the two individual trumpet geometries. We show that these trumpets admit a combinatorial description, which are similar to the `annular half-permutations" appearing in \cite{kusalik_orthogonal_2005, mingo_noncrossing_2019, cebron_infinitesimal_2025}.

An annular non-crossing permutation with $k$ through-blocks can be cut into two non-crossing half-permutations drawn on a disk, each with $k$ open blocks. Conversely, a pair of half-permutations can be reassembled into exactly $k$ annular non-crossing permutations, corresponding to the $k$ possible relative cyclic alignments of their open
blocks. Similar ideas involving cutting and re-gluing annular non-crossing permutations appeared in \cite{cao_ramp_2025}.

\begin{figure}[ht!]
    \centering
    \includegraphics[width=1\linewidth]{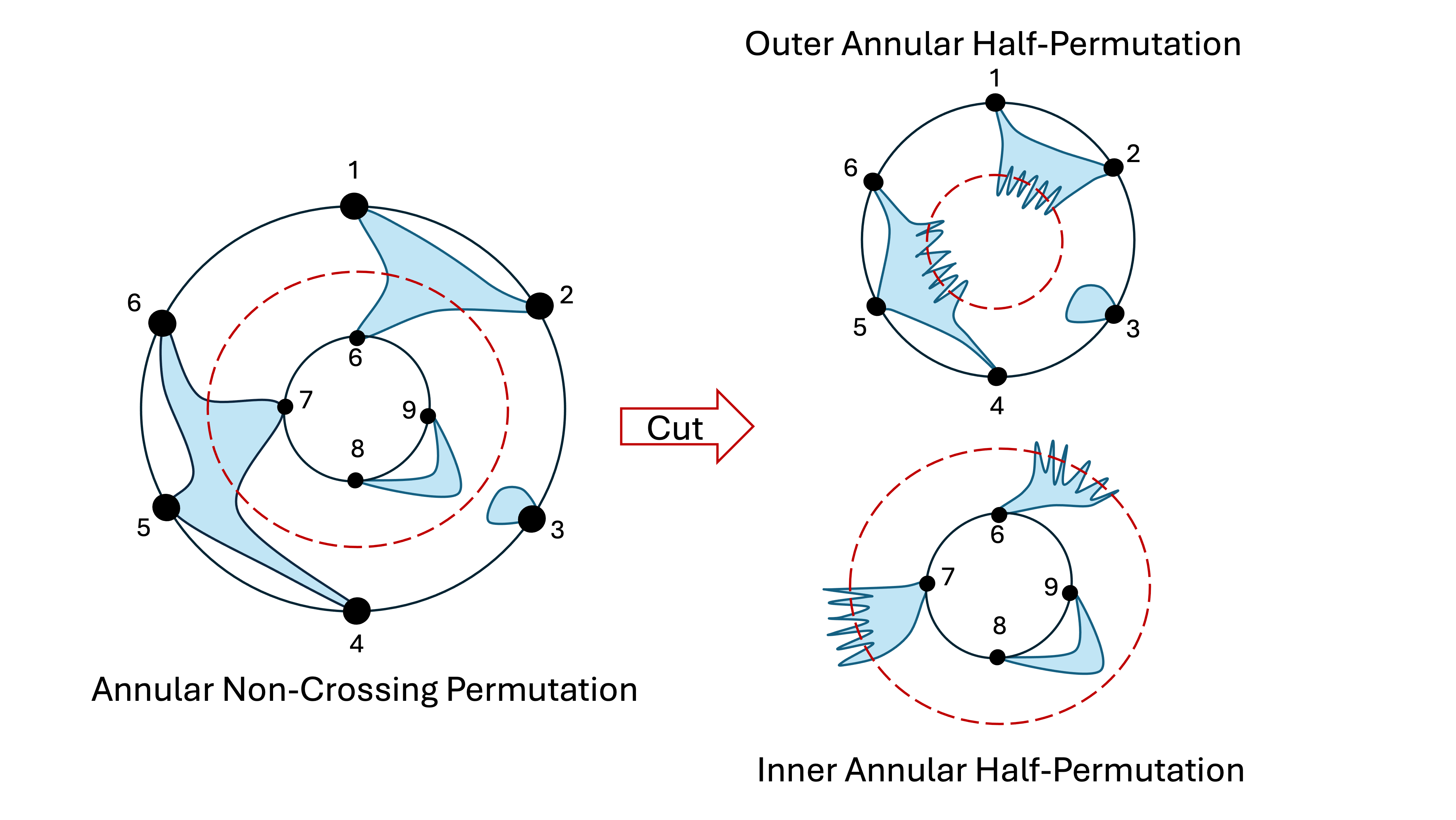}
    \caption{ Cutting an annular non-crossing permutation into two annular half-permutations. The dashed red circle indicates the cut through the $k$ through-connections of the original annular permutation. After cutting, these through-connections become the $k$ open blocks of the outer and inner  $\mathrm{ANC}_{1/2}$ half-permutations, shown by the jagged open ends. Cycles that lie entirely on a single boundary remain closed blocks on the
    corresponding half-permutation. Then, each half is characterized by its number of boundary points, closed blocks, and open blocks, while gluing the two sets of open blocks reconstructs the original annular non-crossing permutation.
    Here, the two open blocks can be re-glued with two possible relative cyclic shifts. More generally, $k$ open blocks admit $k$ such shifts. In the continuum limit, this discrete relative shift becomes the relative twist variable appearing in the JT gravity gluing measure.}
    \label{fig:Half_ANC}
\end{figure}

The number of such annular-half permutations was shown in \cite{kusalik_orthogonal_2005}, which we call ANC$_{1/2}$, to be given by
\begin{equation}
    \#\text{ANC}_{1/2}(n,j,k) = \binom{n}{j}\binom{n}{j+k},
\end{equation}
where $k$ is the number of open blocks, which are the through connections of the original ANC permutation which get cut when making two ANC$_{1/2}(n, j, k)$ permutations, and $j$ is the number of closed blocks drawn on the $n$ boundary points. Closed blocks are cycles that only connect the outer circle to the outer circle or the inner circle and inner circle in the original ANC diagram. Here, the previous through cycle parameter $c$ has been relabeled $k$ for consistency with the mathematics literature and to distinguish its role in these new sets of non-crossing permutations. An illustration for how to cut an ANC permutation into two half-permutations is shown in Fig. \ref{fig:Half_ANC}. Summing over the number of closed blocks yields the number of total annular half permutations with $n$ boundary points and $k$ open blocks, giving

\begin{equation}
    \#\text{ANC}_{1/2}(n, k) = \sum_{j=0}^{n-k} \binom{n}{j}\binom{n}{j+k} =  \binom{2n}{n-k}.
\end{equation}
We immediately recognize the above enumeration as the count in the previous section leading to the trumpet geometry, where the enumeration for annular non-crossing permutations may be written as
\begin{equation}
    \mathcal N(n,m,c)
    =
    c\ \#\text{ANC}_{1/2}(n, c)\ \#\text{ANC}_{1/2}(m, c)\ .
    \label{eqn:half-factorization}
\end{equation}
Here, we change the notation for through cycles back to $c$ to make contact with earlier conventions for ANC permutations. Thus, the two binomial factors may be regarded as the two combinatorial halves of the annulus, while the factor $c$ counts the ways in which they may be sewn together.

\begin{figure}
    \centering
    \includegraphics[width=1\linewidth]{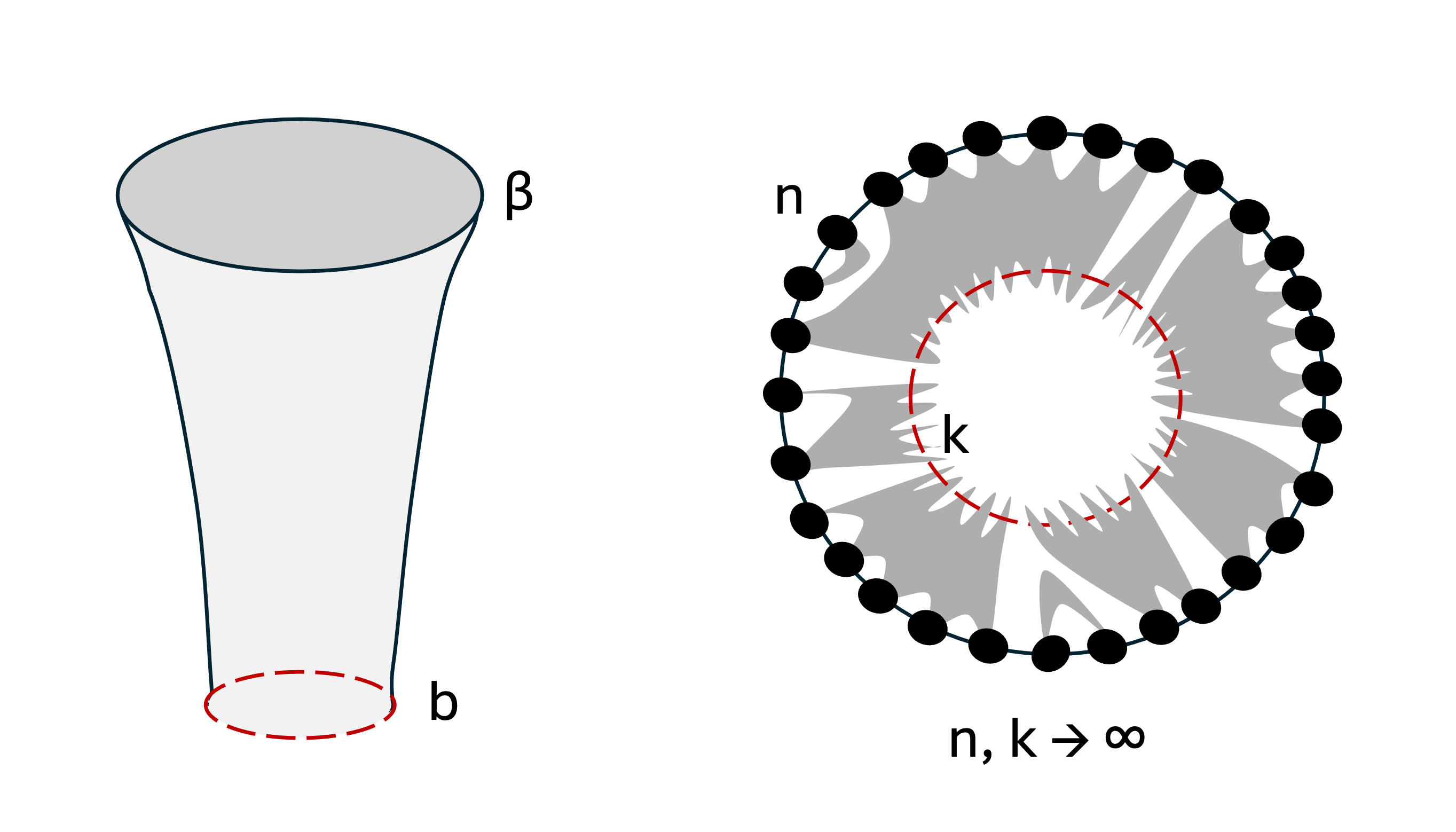}
    \caption{
    Geometric interpretation of an annular half-permutation. An outer circular half-permutation has $n$ marked boundary points and $k$ open blocks exposed along the cut is shown on the right. In the continuum scaling
    $n,k\rightarrow\infty$ with $k/\sqrt{n}$ fixed, the outer boundary is identified with the asymptotic JT boundary of length $\beta$, while the cut
    carrying the $k$ open blocks becomes the geodesic boundary of length $b$, with $n\leftrightarrow\beta$ and $2k\leftrightarrow b$. The resulting continuum geometry is a trumpet, shown on the left. When two such half-permutations are glued, the $k$ possible relative cyclic alignments of their open blocks give the discrete analogue of the double-trumpet gluing
    measure.
    }
    \label{fig:Trumpet_Half_ANC}
\end{figure}

This factorization also has continuum interpretation. In the scaling
$c=O(\sqrt n)$,
\begin{equation}
    4^{-n}\#\text{ANC}_{1/2}(n, c)
    \sim
    \frac{1}{\sqrt{\pi n}}
    e^{-c^2/n}\ .
\end{equation}
With $n\longleftrightarrow\beta$ and $2c\longleftrightarrow b$ this becomes
\begin{equation}
    4^{-n}\#\text{ANC}_{1/2}(n, c)
    \longrightarrow
    \frac{1}{\sqrt{\pi\beta}}
    e^{-b^2/(4\beta)}
    =
    2Z_{\rm tr}(\beta,b)\ .
    \label{eqn:half-trumpet}
\end{equation}
Thus a single annular half-permutation has a continuum limit proportional to the single trumpet partition function in JT gravity. Sewing two such halves and summing over their common number of open blocks then reproduces the double-trumpet gluing integral. Here, the $n$ marked points on the outer boundary become the asymptotic boundary length $\beta$, while the $k$ open blocks produced by cutting the annulus
become the geodesic boundary of the trumpet. This is illustrated in Fig. \ref{fig:Trumpet_Half_ANC}

We will show in later sections how this trumpet enumeration leads to the higher-boundary partition functions, for example, the three boundary pair of pants geometry, suggesting that more general JT geometries derived with topological recursion and trumpet gluing can be described via a collection of annular half-permutations ``sewn" together, forming a higher-boundary non-crossing permutation.

\section{Orientation-Reversing Permutations and Non-Orientable JT Gravity}
\label{nonorientable}

The previous sections showed that continuous geometry can emerge from the parameters appearing in enumerations of non-crossing permutations. We now consider a different type of information carried by the diagrams and permutations. In addition to these continuous parameters, annular non-crossing permutations can possess a discrete symmetry associated with reversing the cyclic orientation of the inner boundary. We will show that this symmetry also has an analogue in the continuum limit. 

This provides an example where the permutation reflects not only topology, but also the geometry of a particular surface and its symmetries. The relative orientation of the boundary cycles distinguishes two classes of annular diagrams that have identical values of $n$, $m$, and $c$, but cannot be related while preserving the cyclic orientation assigned to both boundaries. The two classes are related by a $\mathbb{Z}_2$ reversal symmetry. In the corresponding boundary theory, this distinction is associated with the presence or absence of time-reversal symmetry, while in the bulk theory it is associated with extending the path integral to include orientation reversal and non-orientable geometries \cite{stanford_jt_2020}.

We first describe this symmetry directly by its permutation conditions and the corresponding annular diagrams. We then show that the two sectors have identical enumerations at fixed through-cycle number. Their continuum limits therefore produce two identical copies of the double-trumpet integrand, giving the orientation-reversal multiplicity found in the double-trumpet in JT gravity with non-orientable manifolds.

\subsection{Orientation Reversal of Annular Permutations}

To cast orientation-reversal in permutation language, we review the formalism first introduced in \cite{redelmeier_real_2012}. For a finite set $I$ of positive integers, define
\begin{equation}
    -I=\{-k:k\in I\}\ ,
    \qquad
    \pm I=I\cup(-I)\ ,
\end{equation}
and let $\delta:k\longmapsto -k$ exchange the two copies. The map $\delta$ sends each labeled point to its orientation-reversed counterpart, switching the sign used to distinguish the two orientations. The permutations on $\pm I$ which are premaps are relevant here. A permutation $\pi\in S(\pm I)$ is a premap if
\begin{equation}
    \pi(k)
    =
    -\pi^{-1}(-k)\ ,
    \label{eqn:premap-condition}
\end{equation}
for every $k\in\pm I$, and no cycle of $\pi$ contains both $k$ and $-k$. The premap condition means that if a cycle appears with one orientation, the corresponding cycle with all signs and the cyclic order reversed must also appear. Equivalently,
\begin{equation}
    \delta\pi\delta=\pi^{-1}\ .
    \label{eqn:premap-invariance}
\end{equation}
We denote the set of such permutations by ${\rm PM}(\pm I)$. Equation~\eqref{eqn:premap-invariance} highlights how this permutation condition encodes orientation reversal symmetry. If
\begin{equation*}
    (i_1,i_2,\cdots,i_r)
\end{equation*}
is a cycle of a premap, then there is a distinct companion cycle
\begin{equation}
    (-i_r,\cdots,-i_2,-i_1)\ .
    \label{eqn:premap-cycle-pair}
\end{equation}
In other words, passing from the positive to the negative labels keeps the
same connections but reverses the order in which the points are encountered
around the cycle. In \cite{redelmeier_real_2012}, this negative labeling
keeps track of the opposite orientation of the inner boundary of the annular diagram. This can also be viewed as looking at one of the boundary circles from the opposite side, which reverses its cyclic ordering relative to the other
boundary.

\begin{figure}
    \centering
    \includegraphics[width=1.\linewidth,
    trim=1cm 4.5cm 1cm 3.5cm,
    clip]{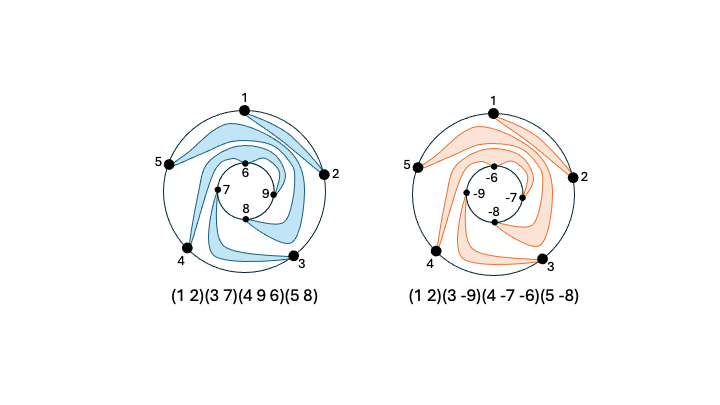}
    \caption{
    The two relative orientations of annular non-crossing premaps. (Left) An $\varepsilon=+1$ component on $V_1\cup V_2$, illustrated by
    $(12)(37)(496)(58)$. The premap condition supplies a partner component on $(-V_1)\cup(-V_2)$ with the cyclic ordering reversed.
    (Right) An $\varepsilon=-1$ component on $V_1\cup(-V_2)$, shown here as $(12)(3{-9})(4{-7}{-6})(5{-8})$; its partner lies on
    $(-V_1)\cup V_2$. Simultaneously reversing the signs of both boundaries exchanges the two components of the same premap, so only the relative sign between the boundaries is meaningful. This leaves the two possibilities $\varepsilon=+1$ and $\varepsilon=-1$,
    corresponding respectively to an inner boundary with the same orientation and sign, or with reversed cyclic ordering and opposite sign.
    }
    \label{fig:ANC_REV}
\end{figure}

For the annular non-crossing permutations considered here, let
\begin{equation*}
    V_1=\{1,\ldots,n\}\ ,
    \qquad
    V_2=\{n+1,\ldots,n+m\}\ ,
\end{equation*}
be the set of labels on the inner and out annulus circle, respectively, with boundary permutation
\begin{equation}
    \gamma
    =
    (1,2,\cdots,n)
    (n+1,n+2,\cdots,n+m)\ .
    \label{eqn:redelmeier-gamma}
\end{equation}
There are two choices for the relative orientation of these cycles, $V_1\cup V_2$ and $V_1\cup(-V_2)$, which can also be labeled by a sign $\varepsilon=\pm1$. For $\varepsilon=+1$, both boundary cycles are represented by their
positively labeled copies. For $\varepsilon=-1$, the second boundary is represented by its negatively labeled copy, and its cyclic ordering reversed relative to the first. Equivalently, the two relative orientations are described by the boundary permutations, $\gamma$ and 
\begin{equation}
    \gamma_{\rm op}
    =
    (1,2,\cdots,n)
    (-n-m,\cdots,-n-2,-n-1)\ ,
    \label{eqn:gamma-op}
\end{equation}
where $\gamma_{\rm op}$ denotes the opposite relative orientation of the two circles \cite{redelmeier_real_2012}.
These two classes of permutations lead to two sets of annular non-crossing permutations, denoted $S_{\rm ann-nc}(\gamma)$ and $S_{\rm ann-nc}(\gamma_{\rm op})$.  A comparison of these two classes of annular non-crosing permutations is illustrated in Fig.~\ref{fig:ANC_REV}. 


The important point here is the discrete orientation symmetry is encoded in the permutation conditions. The quantities $n$, $m$, and $c$ measure the size and connectivity of an annular diagram, while $\varepsilon$ distinguishes its two possible relative boundary orientations. Changing $\varepsilon$ leaves the numbers of marked points and through-cycles unchanged. We can characterize the annular permutations informally as  the set of parameters $(n,m,c,\varepsilon)$ with $\varepsilon=\pm1$ carrying information that is independent of the enumerative parameters that will be scaled in the continuum limit but which nevertheless carries physical information about the symmetry of the emergent geometry.

\subsection{Enumeration of the Two Orientations and the Continuum Limit}

Since the two sets of ANC permutations differ only by reversing the relative orientation of the inner boundary of the annulus, the number of total diagrams for both are the same, and so for $\lambda=1$, both are given by the same enumeration considered in the previous section, giving
\begin{equation}
    \mathcal{N}_{\gamma}(n,m,c)
    =
    \mathcal{N}_{\gamma_{\rm op}}(n,m,c)  =
    c
    \binom{2n}{n-c}
    \binom{2m}{m-c}\ .
    \label{eqn:orientation-count-equality}
\end{equation}
The total permutation count includes both relative orientations and gives
\begin{equation}
    \mathcal{N}_{\pm}(n,m,c)
    =
    \mathcal{N}_{\gamma}(n,m,c)
    +
    \mathcal{N}_{\gamma_{\rm op}}(n,m,c)
    =
    2c
    \binom{2n}{n-c}
    \binom{2m}{m-c}\ .
    \label{eqn:doubled-annular-count}
\end{equation}
We highlight that this factor of two appears before any continuum limit is taken and, moreover, has a different combinatorial origin from the factor $c$.\footnote{The same factor of two appears in the spectral form factor. In random matrix theory, the early ramp in the orthogonal class ($\beta_{\rm D}=1$) has twice the slope of the unitary class ($\beta_{\rm D}=2$) \cite{mehta_random_1967,forrester_review_2022}.
This agrees with two cases of the ramp in JT gravity, with and without non-orientable surfaces included \cite{weber_unorientable_2024}.} The latter counts relative cyclic alignments of the through-cycles and eventually becomes part of the continuous gluing measure. The factor of two counts the two possible relative orientations of the two boundary cycles and
reflects the symmetry of the permutations considered here.

Since changing the relative orientation leaves $n$, $m$, and $c$ unchanged, the two sectors have identical continuum limits. Applying the same procedure as in the previous section gives
\begin{equation}
    \mathcal{N}_{\pm}(n,m,c)
    \sim
    2
    \frac{4^{n+m}}{\pi\sqrt{nm}}
    c
    e^{-c^2/n}
    e^{-c^2/m}\ .
    \label{eqn:doubled-annular-continuum}
\end{equation}
The orientation information therefore factors completely from the dependence
on the scaling variables.

\begin{figure}
    \centering
    \includegraphics[width=1\linewidth,
    trim=0.5cm 2.5cm 0.5cm 1.5cm,
    clip]{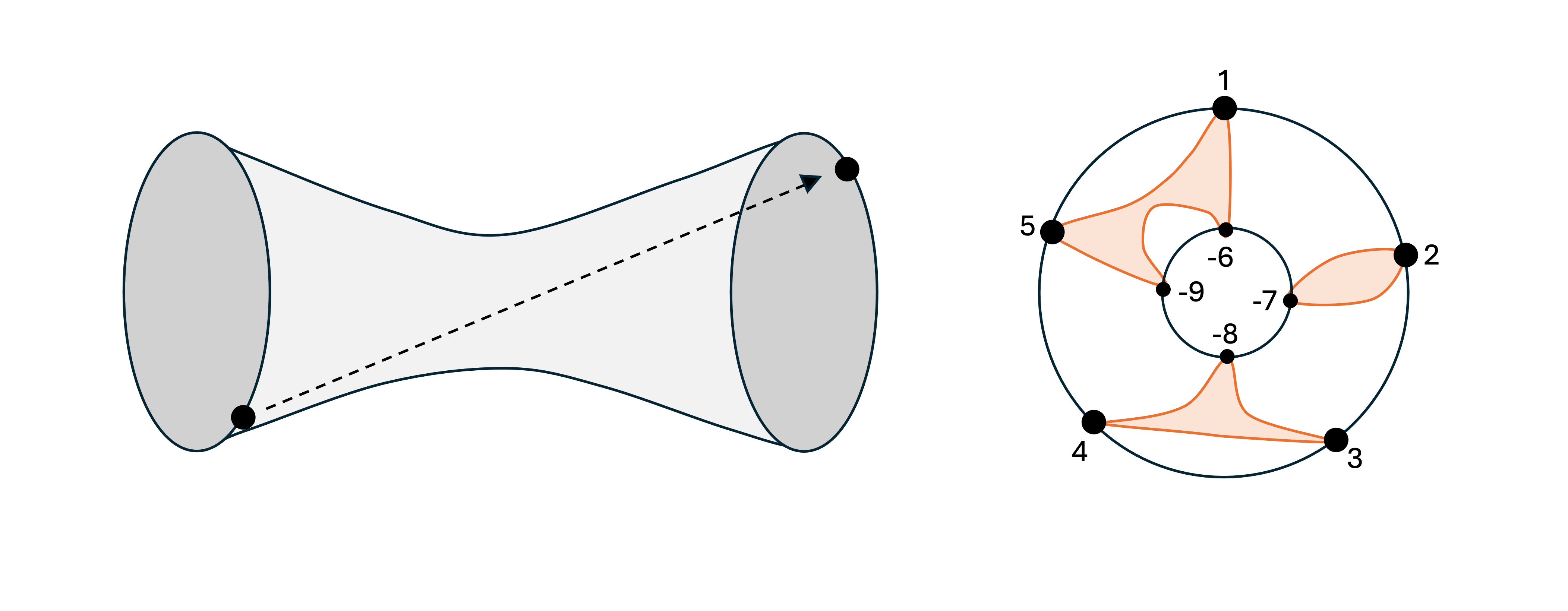}
    \caption{
    (Left) Orientation-reversing identification of two trumpets. Following the
    dashed path through the identification reverses the local orientation
    \cite{yan_crosscap_2022}.
    (Right) The corresponding reversed annular non-crossing diagram, with the
    inner boundary cyclic ordering reversed. In the continuum limit of the number of points on both boundaries and the number of through connections, this diagram reproduces the geometry shown to the left.
    }
    \label{fig:dt-reversed}
\end{figure}

Replacing the sum over $c$ by an integral gives
\begin{equation}
    \sum_c\mathcal{N}_{\pm}(n,m,c)
    \sim
    2 \times
    \frac{4^{n+m}}{\pi\sqrt{nm}}
    \int_0^\infty
    c\ dc
    e^{-c^2/n}
    e^{-c^2/m}\ .
    \label{eqn:doubled-annular-integral}
\end{equation}
Using the dictionary found in the previous section and removing the nonuniversal edge factor, this becomes
\begin{equation}
    2
    \int_0^\infty
    b\ db
    Z_{\rm tr}(\beta_1,b)
    Z_{\rm tr}(\beta_2,b)\ .
    \label{eqn:doubled-double-trumpet}
\end{equation}
Following \cite{stanford_jt_2020}, the double-trumpet partition function can be written as
\begin{equation}
    Z_{\rm DT}(\beta_1,\beta_2)
    =
    c_{\rm SW}
    \int_0^\infty
    bdb
    Z_{\rm tr}(\beta_1,b)
    Z_{\rm tr}(\beta_2,b)\ ,
    \label{eqn:double-trumpet-sw}
\end{equation}
where $Z_{\rm tr}(\beta,b)$ is the trumpet partition function and $c_{\rm SW}$ is a factor accounting for the allowed orientation choices in the gluing. When orientation reversal is not included, $c_{\rm SW}=1$. If orientation reversal is included, the two boundaries may be glued with either relative orientation, giving $c_{\rm SW}=2.$ Thus, we see that the factor of two from the annular non-crossing permutations indeed captures this symmetry. The orientation-reversing diagrams are compared to the double trumpet in Fig. \ref{fig:dt-reversed}.


\section{Higher-Boundary Planar Permutations}
\label{sec:higher-order-jt}

The relationship between the disk, annulus, and their counterparts in JT gravity suggests a more general mapping between non-crossing permutations drawn on multiple boundaries and multi-boundary JT geometries.
The higher-boundary case also allows us to separate two pieces of this correspondence. Previous work of Do, He, and Mathews related the leading terms of \textit{pruned }non-crossing enumerations to intersection numbers and to the top-degree part of Weil--Petersson volumes
\cite{do2019countingnoncrossingpermutationssurfaces}. These pruned enumerations are formally similar to those considered throughout this paper. However, this pruning is the process of stripping away the binomials which scale to the trumpet geometries in the continuum limit studied in the previous section. Here, we study the continuum scaling of the full boundary-dependent enumeration and compare it directly with the JT partition function obtained after attaching
trumpets to the Weil--Petersson core, computing the 3 and four boundary cases as examples. For three boundaries this reproduces the full genus-zero partition function since $V_{0,3}=1$, while for four or more
boundaries the construction continues to capture only the top-degree Weil--Petersson volume with trumpets attached. To conclude, we also test our construction against a higher genus enumeration and compare our results to the analogous geometry in JT gravity.

\subsection{Enumeration of Multi-Boundary Non-crossing Permutations.}

Non-crossing permutations on multi-boundary planar surfaces have a rich structure, and have been extensively studied  \cite{collins_second_2006}. Here, we will focus only on their enumeration so that we can test the generality of the continuum limit and its relation to geometries in JT gravity. Let
\begin{equation*}
    L=\sum_{i=1}^{q}n_i
\end{equation*}
be the total number of marked points on all $q$ boundaries, and let
\begin{equation*}
    \gamma_{\mathbf n}
    =
    (1,\cdots,n_1)
    (n_1+1,\cdots,n_1+n_2)\cdots
\end{equation*}
encode their cyclic ordering on the $q$ boundaries.
A permutation $\pi\in S_L$ defines a connected planar diagram when when all $q$ boundaries are connected and it satisfies a generalization of the genus-zero Euler relation. The number of such non-crossing permutations is \cite{collins_second_2006}
\begin{equation}
    2
    \frac{(L-1)!}{(L-q+2)!}
    \prod_{i=1}^{q}
    \left[
        n_i\binom{2n_i-1}{n_i}
    \right]\ .
    \label{eq:cmss-count}
\end{equation}
This formula reduces to the Catalan numbers for $q=1$ and to the number of annular non-crossing permutations for $q=2$, and so is the multi-boundary extension of these enumerations.

\subsection{Continuum Limit and Top-Degree Terms in JT Gravity}

For fixed $q$ and large $n_i$,
\begin{equation}
    n_i\binom{2n_i-1}{n_i}
    \sim
    \frac{4^{n_i}}{2\sqrt{\pi}}\sqrt{n_i}\ ,
\end{equation}
and
\begin{equation}
    \frac{(L-1)!}{(L-q+2)!}
    \sim
    L^{q-3}.
\end{equation}
Equation~\eqref{eq:cmss-count} gives
\begin{equation}
    2
    \frac{(L-1)!}{(L-q+2)!}
    \prod_{i=1}^{q}
    \left[
        n_i\binom{2n_i-1}{n_i}
    \right]
    \sim
    \frac{2^{1-q}}{\pi^{q/2}}
    L^{q-3}
    \prod_{i=1}^{q}\sqrt{n_i}\ .
    \label{eq:cmss-asymptotic}
\end{equation}
The factor of $4^L$ was normalized out in the right-hand-side. 

The appearance of one large-$n_i$ boundary factor for each boundary is consistent with the trumpet interpretation developed in the case of annular non-crossing permutations. There, resolving the enumeration before performing the gluing sum showed that a half-permutation becomes a single JT trumpet in the continuum limit. In the present higher-boundary enumeration the internal connection data have already been summed over, so the individual geodesic lengths $b_i$ are no longer resolved. Nevertheless, the resulting asymptotic count can be compared directly with the JT quantity obtained after integrating the corresponding Weil--Petersson volume against one trumpet for each boundary.

On the JT gravity side, the highest-degree terms of the Weil--Petersson volume polynomial are determined by $\psi$-class intersection numbers \cite{Mirzakhani:2006eta, mertens_solvable_2023}. In general, the coefficient of $\prod_i b_i^{2d_i}$ in the top-degree part is proportional to
$\langle\tau_{d_1}\cdots\tau_{d_q}\rangle_0$. At genus zero these intersection numbers are \cite{Johnson:2026wlo}
\begin{equation*}
    \left\langle
        \tau_{d_1}\cdots\tau_{d_q}
    \right\rangle_0 
    = \int_{\overline{\mathcal M}_{0,q}} \prod_{i=1}^q \psi_i^{d_i}
    =
    \frac{(q-3)!}{\prod_i d_i!}\ ,
\end{equation*}
where $\tau_{d_i}$ denotes the standard $\psi$-class insertion of degree $d_i$ associated with the $i$th boundary, and
$\sum_i d_i=q-3$, which gives
\begin{equation}
    V_{0,q}^{\rm top}(b_1,\ldots,b_q)
    =
    \frac{1}{2^{q-3}}
    \sum_{\substack{d_1+\cdots+d_q=q-3}}
    \frac{(q-3)!}{\prod_i(d_i!)^2}
    \prod_{i=1}^{q}b_i^{2d_i} \, ,
    \qquad q\geq3\ .
    \label{eq:wp-top}
\end{equation}
To pass from fixed geodesic boundaries to asymptotic boundaries, we attach a trumpet to each boundary and integrate over its geodesic length, giving
\begin{equation}
    \int_0^\infty
    b\ db\ b^{2d}Z_{\rm tr}(\beta,b)
    =
    \frac{2^{2d}d!}{\sqrt{\pi}}
    \beta^{d+1/2}\ .
    \label{eq:trumpet-moment}
\end{equation}
Substituting this into Eq.~\eqref{eq:wp-top}, the resulting JT partition function is
\begin{equation}
    \widehat Z_{0,q}^{\rm JT,top}
    (\beta_1,\ldots,\beta_q)
    =
    \frac{2^{q-3}}{\pi^{q/2}}
    \left(
        \sum_{i=1}^{q}\beta_i
    \right)^{q-3}
    \prod_{i=1}^{q}\sqrt{\beta_i}\ .
    \label{eq:jt-top-q}
\end{equation}

The ratio between the scaled enumeration and its JT counterpart at fixed boundary number gives the factor $4^{2-q}$, which a normalization constant depending on the number of boundaries. After normalization, the continuum limit of the higher-boundary enumeration reproduces the result of integrating $V_{0,q}^{\rm top}$ against $q$ trumpet partition functions. The appearance of the top-degree Weil--Petersson volume is consistent
with \cite{do2019countingnoncrossingpermutationssurfaces}. Here, we extend their results to JT gravity by showing that the large-boundary scaling limit of the enumeration yields the corresponding
JT partition function with arbitrary number of asymptotic boundaries. In the annular case, the boundary data decomposed into combinatorial
half-permutations whose continuum limits are JT trumpets. In this higher-boundary formula, that intermediate gluing form has already been summed over. We end this section by computing the first few continuum limits of multi-boundary non-crossing enumerations showing they limit to the top-degree parts of the partition functions for the pair of pants with three asymptotic boundaries and the four boundary geometry in JT gravity. These are also the first stable geometries in topological recursion.

\subsection{Higher-Boundary Examples}

We first consider three boundaries. In this case
Eq.~\eqref{eq:cmss-count} reduces to
\begin{equation*}
    \mathcal{C}_{0,3}(n_1,n_2,n_3)
    =
    2
    \prod_{i=1}^{3}
    \left[
        n_i\binom{2n_i-1}{n_i}
    \right]\ ,
\end{equation*}
where $\mathcal{C}_{0,3}$ denotes the corresponding non-crossing
permutation count. Its large-boundary behavior is
\begin{equation}
    4^{1-L}
    \mathcal{C}_{0,3}(n_1,n_2,n_3)
    \sim
    \frac{1}{\pi^{3/2}}
    \sqrt{n_1n_2n_3}\ .
    \label{eq:three-boundary-count}
\end{equation}

The corresponding genus-zero three boundary geometry in JT gravity is the pair of pants. This case is
particularly simple because the pair-of-pants moduli space is
zero-dimensional and
\begin{equation*}
    V_{0,3}(b_1,b_2,b_3)=1\ .
\end{equation*}
Consequently, the nontrivial dependence on the asymptotic boundary lengths
comes entirely from the three trumpet factors attached to the geodesic
boundaries of the pair-of-pants core. After performing the three trumpet
integrals one obtains
\begin{equation}
    \widehat Z_{0,3}^{\rm JT}
    =
    \frac{1}{\pi^{3/2}}
    \sqrt{\beta_1\beta_2\beta_3}\ .
    \label{eq:three-boundary-jt}
\end{equation}
The hat on $Z$ reflects the fact that we are only considering the top-degree part of the underlying Weil-Petersson volumes, though this distinction is trivial for three boundaries since $V_{0,3} =1$. Under $n_i\leftrightarrow\beta_i$, the large-boundary scaling in
Eq.~\eqref{eq:three-boundary-count} reproduces the full genus-zero three-boundary JT result, up to normalization. The additional $\sqrt{\beta_1\beta_2\beta_3}$ dependence appearing here is the term generated by attaching one trumpet to each of the three geodesic boundaries. The three-boundary case therefore provides the simplest example in which the continuum scaling of the full enumeration may be
compared directly with the trumpet-dressed geometries appearing in JT gravity.

\begin{figure}
    \centering
    \includegraphics[width=1.\linewidth,
    trim=1cm 2.cm 1cm 2.cm,
    clip]{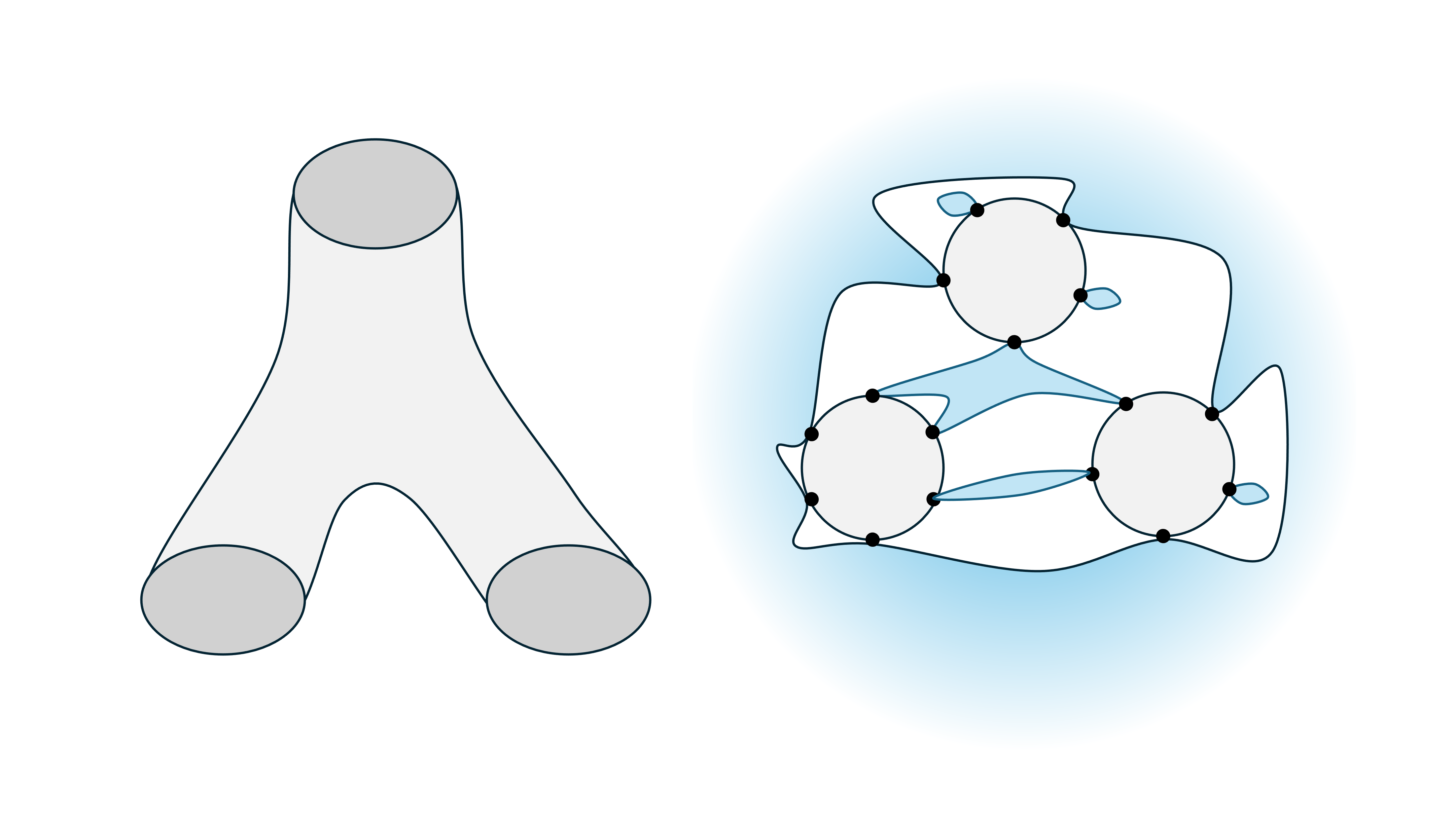}
    \caption{
    Three-boundary non-crossing diagrams and the pair-of-pants geometry.
    (Left) A genus-zero surface with three boundaries is a pair of pants.
    (Right) A connected non-crossing permutation drawn on three boundary
    circles. The exterior region represents the connected genus-zero surface
    obtained by removing three disks from the sphere.
    }
    \label{fig:Pants}
\end{figure}

Geometrically, the corresponding non-crossing diagrams have the topology of
a pair of pants, as illustrated in Fig.~\ref{fig:Pants}. The pair of pants is
the basic three-boundary building block in the pants decomposition of hyperbolic surfaces used in JT gravity and topological recursion \cite{saad_jt_2019,mertens_solvable_2023}. The same topology also appears in string theory as the worldsheet describing the splitting or joining of three closed strings \cite{kiritsis_string_2019}.

The three-boundary example, however, cannot distinguish between recovering
the top-degree Weil--Petersson data and recovering the full Weil--Petersson
volume, because for $V_{0,3}$ the two are identical. The first nontrivial
test therefore occurs at four boundaries. In this case the moduli space has
positive dimension and the Weil--Petersson volume contains both a
highest-degree term and a lower-degree contribution,
\begin{equation}
    V_{0,4}(b_1,b_2,b_3,b_4)
    =
    2\pi^2
    +
    \frac{1}{2}\sum_{i=1}^{4}b_i^2 \ .
    \label{eq:wp-four}
\end{equation}

\begin{figure}[t]
    \centering
    \includegraphics[width=0.7\linewidth]{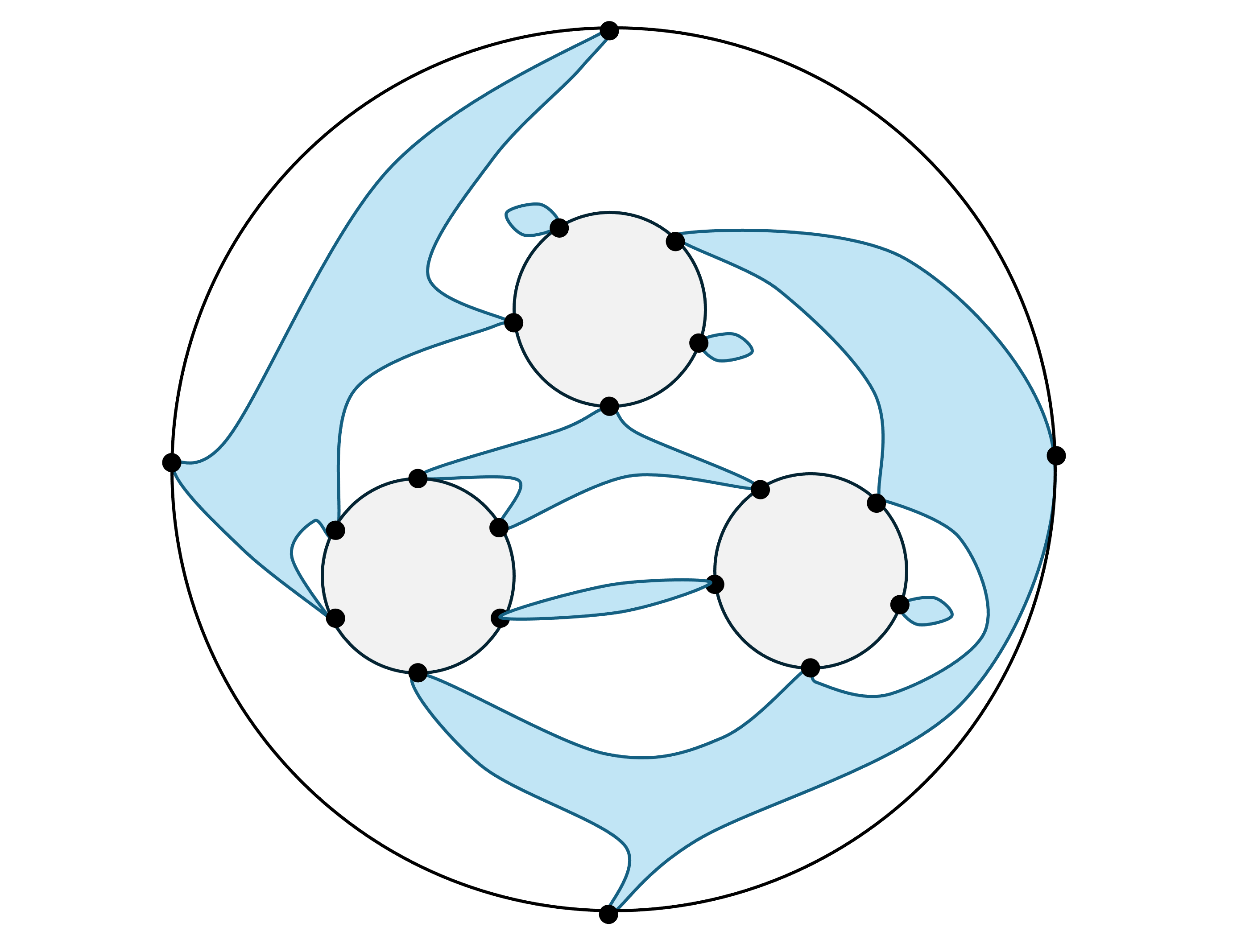}
    \caption{
    Example of a connected genus-zero non-crossing permutation drawn on four boundary components. Here, the diagram includes one cycle that connects all four boundaries. This is an alternative way to draw multi-boundary non-crossing permutations. In the continuum description the corresponding hyperbolic surface has a nontrivial moduli space, whose Weil--Petersson volume is given by Eq.~\eqref{eq:wp-four}.
    }
    \label{fig:FourBoundary}
\end{figure}

The corresponding four-boundary non-crossing diagrams are again connected genus-zero objects, now drawn on four boundary components, as illustrated in  Fig.~\ref{fig:FourBoundary}. Unlike the pair of pants, a four-boundary sphere possesses a nontrivial modulus, and its
Weil--Petersson volume consequently depends on the boundary lengths, as shown in Eq.~\eqref{eq:wp-four}.

For $q=4$, Eq.~\eqref{eq:cmss-count} gives
\begin{equation}
    \mathcal{C}_{0,4}(n_1,n_2,n_3,n_4)
    =
    2(L-1)
    \prod_{i=1}^{4}
    \left[
        n_i\binom{2n_i-1}{n_i}
    \right],
\end{equation}
whose large-boundary behavior is, after normalization, 
\begin{equation}
    \mathcal{C}_{0,4}(n_1,n_2,n_3,n_4)
    \sim
    \frac{2}{\pi^2}
    \left(\sum_{i=1}^{4}n_i\right)
    \prod_{i=1}^{4}\sqrt{n_i}\ .
    \label{eq:four-boundary-count}
\end{equation}
After attaching four trumpets, the resulting JT partition function is
\begin{equation}
    \widehat Z_{0,4}^{\rm JT}
    =
    \frac{2}{\pi^2}
    \prod_{i=1}^{4}\sqrt{\beta_i}
    \left(
        \sum_{i=1}^{4}\beta_i+\pi^2
    \right)\ .
    \label{eq:jt-four}
\end{equation}
Under $n_i\leftrightarrow\beta_i$, Eq.~\eqref{eq:four-boundary-count}
reproduces precisely the term proportional to
$\sum_i\beta_i$. This is the contribution obtained by attaching trumpets to
the highest-degree part
\begin{equation*}
    V_{0,4}^{\rm top}
    =
    \frac{1}{2}\sum_{i=1}^{4}b_i^2\ ,
\end{equation*}
but it does not reproduce the contribution arising from the constant
$2\pi^2$ term in Eq.~\eqref{eq:wp-four}.

The four-boundary example highlights the limitation of the present correspondence between topologies and geometries in JT gravity and non-crossing permutations. The continuum scaling supplies the boundary dependence associated with the trumpet dressing of the top-degree Weil--Petersson volume, but the trumpet factors do not generate the lower-degree information missing from the underlying non-crossing enumeration. This is consistent with the result of
\cite{do2019countingnoncrossingpermutationssurfaces} that the leading enumerative data are governed by the top-degree Weil--Petersson terms. Information beyond this leading sector must therefore enter through additional combinatorial data, corresponding on the matrix-model side to information beyond the universal spectral-edge behavior. 

\subsection{A Higher-Genus Example}

Although the preceding analysis has focused on genus-zero geometries, we can test the same continuum scaling in the simplest higher-genus case. This example plays a different role from the disk, annulus, and higher-boundary planar geometries considered above, which arise directly
from the expansion of matrix moments and connected
multi-trace correlators in terms of planar, non-crossing diagrams \cite{mingo_annular_2004,mingo_second_2007,collins_second_2006}. A systematic treatment of nonzero genus would require a genus-refined extension of this framework, such as those developed in \cite{borot_functional_2023}.\footnote{Appendix~\ref{app:resolvent-rtransform} reviews the
free probability/matrix model framework underlying the planar and multi-boundary cases, including the roles of resolvents, the $R$-transform, and higher-order free probability. The higher-genus polygon enumerations used here, however, are not derived from the free cumulant expansions appearing in those frameworks.} Instead, we use the $(g,n)=(1,1)$ enumeration of
\cite{do2019countingnoncrossingpermutationssurfaces} as a first check that the continuum scaling of non-crossing diagrams persists beyond the planar setting. More generally, that work considers non-crossing polygon diagrams on surfaces of arbitrary genus and derives recursions whose leading coefficients are
governed by intersection numbers, reproducing the top-degree
Weil--Petersson data. Here, we consider only the resulting enumerations and their continuum scaling, rather than deriving a systematic treatment of all possible diagrams. We refer the reader to their work for more information and illustrations of the diagrams and combinatorics used below.

For the once-punctured torus, let $\nu$ denote the number of marked points on the single boundary of the pruned diagram. The enumeration of such diagrams is
\begin{equation*}
Q_{1,1}(\nu)
=
\begin{cases}
\dfrac{\nu^3-\nu}{24},
& \nu>0\ \text{odd},\\[6pt]
\dfrac{\nu^3+8\nu}{24},
& \nu>0\ \text{even},
\end{cases}
\end{equation*}
so the leading term is independent of whether $\nu$ is even or odd,
\begin{equation}
    Q_{1,1}^{\rm top}(\nu)
    =
    \frac{\nu^3}{24}\ .
    \label{eq:q11-top}
\end{equation}
The coefficient $1/24$ is the genus-one intersection number
$\langle\tau_1\rangle_1$, and is the same intersection data appearing in the highest-degree part of the Weil--Petersson volume.

\begin{figure}[t!]
    \centering
    \includegraphics[width=1\linewidth]{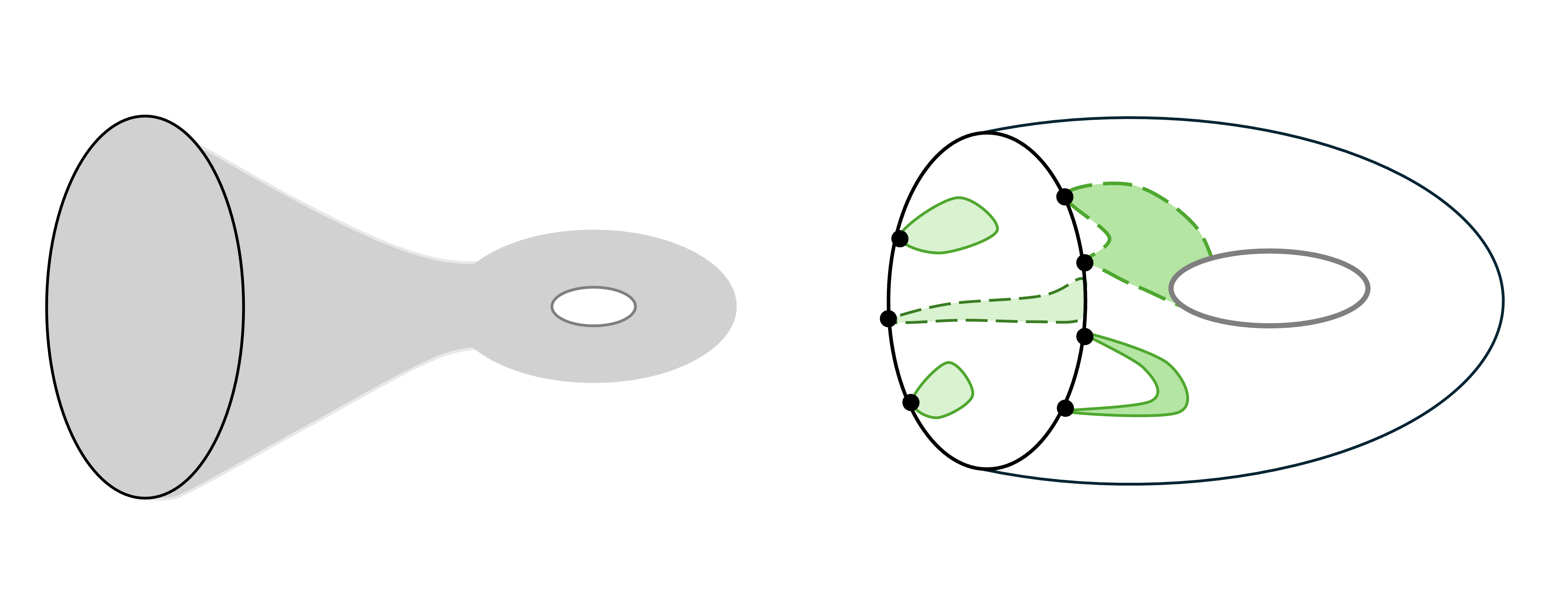}
    \caption{
    A genus-one non-crossing polygon diagram on a surface with one boundary.
    (Left) The corresponding genus-one, one-boundary geometry, which is a  torus with one puncture.
    (Right) An example non-crossing diagram with marked points on the boundary. Some cycles remain localized near the boundary, while others traverse the handle and probe the nontrivial topology of the surface. Dashed segments indicate portions of cycles that pass through the handle
    and are hidden from view.
    }
    \label{fig:genus-one-nc}
\end{figure}

The full unpruned diagrams allow boundary-parallel edges. Let $P_{1,1}(\mu)$ denote their enumeration when the original boundary contains $\mu$ marked points. The topology of the corresponding higher-genus diagrams is illustrated in Fig.~\ref{fig:genus-one-nc}.\footnote{Similar higher-genus permutations appear in \cite{Hock2024}} In contrast with the planar examples considered above, non-crossing cycles on the once-punctured torus may extend through the handle while remaining non-crossing on the surface. Pruning such a diagram leaves a core with $\nu$ marked boundary points, while the removed boundary-parallel data can be restored combinatorially. The relation between the two enumerations is
\begin{equation}
    P_{1,1}(\mu)
    =
    \sum_{\nu=0}^{\mu}
    Q_{1,1}(\nu)
    \binom{2\mu}{\mu-\nu}.
    \label{eq:p11-unpruning}
\end{equation}
Then, $\mu$ measures the size of the full boundary, while $\nu$ measures the boundary of the pruned combinatorial core. Importantly, the term that restores the pruned diagram is the same displaced central binomial coefficient encountered in the half-permutation description of the trumpet.
We can take the same continuum limit as before. For
$\mu\rightarrow\infty$ with $\nu=O(\sqrt{\mu})$,
\begin{equation*}
    4^{-\mu}
    \binom{2\mu}{\mu-\nu}
    \sim
    \frac{1}{\sqrt{\pi\mu}}
    e^{-\nu^2/\mu}\ .
\end{equation*}
With the identifications
\begin{equation}
    \mu\longleftrightarrow\beta\ ,
    \qquad
    2\nu\longleftrightarrow b\ ,
\end{equation}
the unpruning kernel becomes
\begin{equation}
    4^{-\mu}
    \binom{2\mu}{\mu-\nu}
    \longrightarrow
    2Z_{\rm tr}(\beta,b)\ .
\end{equation}
In this form, the full, unpruned, form of the polygon diagram becomes the asymptotic JT boundary, while the boundary of the pruned core becomes the geodesic along which the trumpet is attached. Interestingly, Eq.~\eqref{eq:p11-unpruning} shows that the binomial transform relating the core and full polygon enumerations in \cite{do2019countingnoncrossingpermutationssurfaces} has a direct geometric interpretation in the continuum limit, where each displaced binomial becomes a JT trumpet, so that the discrete transform becomes the operation of attaching asymptotic boundaries to the corresponding Weil--Petersson geometry.

Keeping only the leading part of the pruned enumeration and applying a boundary-dependent
normalization used above, the continuum limit gives
\begin{equation}
    4^{2-\mu} P_{1,1}(\mu)
    \sim
    4\int_0^\infty d\nu\,
    \frac{\nu^3}{24\sqrt{\pi\mu}}\,
    e^{-\nu^2/\mu}
    =
    \int_0^\infty b\,db\,
    \frac{b^2}{48}\,
    Z_{\rm tr}(\beta,b).
    \label{eqn:genus-one-continuum}
\end{equation}
The factor multiplying the trumpet is the highest-degree part of the once-punctured-torus Weil--Petersson volume,
\begin{equation*}
    V_{1,1}(b)
    =
    \frac{b^2+4\pi^2}{48}
    \quad\Longrightarrow\quad
    V_{1,1}^{\rm top}(b)
    =
    \frac{b^2}{48}.
\end{equation*}
Hence, after normalization, Eq.~\eqref{eqn:genus-one-continuum} takes resembles the JT geometry obtained by gluing a trumpet
\begin{equation}
    4^{2-\mu} P_{1,1}(\mu)
    \sim
    \int_0^\infty b\,db\,
    V_{1,1}^{\rm top}(b)\,
    Z_{\rm tr}(\beta,b)
    =
    \frac{\beta^{3/2}}{12\sqrt{\pi}}.
    \label{eqn:genus-one-jt}
\end{equation}
As in the four-boundary genus-zero example, the lower-degree term proportional to $\pi^2$ is not captured by this leading scaling.
Equivalently, Do, He, and Mathews give the closed form
\begin{equation}
    P_{1,1}(\mu)
    =
    \binom{2\mu-1}{\mu}
    \frac{\mu^3+3\mu^2+20\mu-12}
         {12(2\mu-1)}\ ,
\end{equation}
whose leading asymptotic reproduces the same result
\cite{do2019countingnoncrossingpermutationssurfaces}.
The derivation above also explains why only the leading term of the Weil--Petersson volume appears. The pruned enumeration supplies the top-degree Weil--Petersson contribution, while the binomial unpruning binomial becomes the JT trumpet in the continuum limit. As in the four-boundary genus-zero example, the lower-degree term proportional to $\pi^2$ is not captured by this leading scaling. This computation thus provides a first check that the trumpet interpretation is not
restricted to genus zero.

Our explicit higher-genus analysis is limited to the $(g,n)=(1,1)$ case. However, the polygon-diagram enumerations of Do, He, and Mathews have the same boundarywise structure for arbitrary genus and number of boundaries
\cite{do2019countingnoncrossingpermutationssurfaces}. Their results therefore suggest that the scaling observed in the genus-one example, and hence the trumpet interpretation, extends to general $(g,n)$.

\section{Discussion}

This paper studied a relationship between the continuum scaling of enumerations of non-crossing permutations and the genus-zero geometries of JT gravity. In the scaling limit, the discrete boundary sizes become asymptotic boundary lengths, while additional enumerative parameters, such as the number of through-connections in an annular diagram, become continuous moduli of the limiting hyperbolic surface. Under certain identifications of parameters, the JT geometries emerge from these continuum limits. The continuum limit non-crossing permutations on the disk was identified with the low temperature limit of the disk geometry in JT gravity.


The annular non-crossing permutations, on the other hand, have enumerations with scaling limits which can be identified with the full double trumpet expression. This is due to the double trumpet being universal. In addition, this case allowed us to show how discrete permutation rules involving the orientation of points on the boundaries could be mapped to orientation-reversal symmetries in the continuum limit, establishing a connection between permutations and physical geometric symmetries. Furthermore, we demonstrated this pattern extends to an arbitrary number of boundaries at genus-0, highlighting that the non-crossing diagrams only captured the top-degree part of the Weil-Petersson volumes appearing in the gravitational partition functions.

Our results relating combinatorics and the emergent geometric and topological structures in JT gravity path-integral amplitudes complement the more familiar spectral descriptions of matrix-model observables, which are typically formulated in terms of resolvents and spectral densities.  
However, there are three main limitations in our present construction.  First, the unit weights assigned to every cycle type in non-crossing diagrams capture only the top-degree part of the Weil--Petersson volumes, and so the lower-degree terms may require additional combinatorial weights, or even generalized classes of diagrams.  Second, for three or more boundaries, the closed-form enumeration used here sums over some of the connection data that one would need in order to exhibit all internal geodesic moduli directly.  A refined enumeration retaining the incidence data at each boundary should provide a higher-boundary analogue of the through-connection number on the annulus. Third, extending the present analysis systematically beyond the planar, genus-zero setting requires a more general treatment of higher-genus permutation combinatorics. While the cases studied here isolate the basic structure of the correspondence, more general topologies may introduce additional combinatorial data and geometric phenomena that are not captured by the present construction.

This work presents a few interesting future directions. One of our main conclusions was that the permutation symmetry manifests as geometric symmetry in the scaling limit. Diagram symmetry was also shown to be related to the boundary theory's random matrix class. It would be interesting to extend this analysis to the symplectic universality class, whose large-$N$ diagrammatics involve additional orientation and pairing data \cite{redelmeier_quaternionic_2015}.  Determining whether these structures admit a continuum interpretation analogous to the one developed here would provide a test of how broadly the permutation--geometric dictionary extends across random-matrix symmetry classes.

A second direction is the connection to SYK, which provides another setting in which large-$N$ physics possesses a rich combinatorial expansion, while
its low-energy dynamics is governed by the Schwarzian theory \cite{maldacena_remarks_2016,maldacena_conformal_2016}. It would therefore be interesting to investigate the relationship between the diagrams studied here and the chord diagrams appearing in double-scaled SYK
\cite{lin_bulk_2022,berkooz_cordial_2024,berkooz_chaos_2024, pluma2020dynamicalversionsykmodel}.

An open question is whether this combinatorial description extends beyond the perturbative genus expansion and admits a microscopic interpretation. The nonperturbative completion of JT gravity contains effects, such as D-branes and instantons, that are invisible at every fixed order in the topological expansion. Identifying discrete structures that capture these effects would help determine whether the combinatorics developed here are intrinsically perturbative or can also
illuminate the nonperturbative completion of the JT gravitational path integral as encoded by the matrix integral.

Finally, the present work combines two related but distinct combinatorial frameworks because they retain different kinds of information. The planar combinatorics arising in free probability is deeply tied to random-matrix moments and
is sensitive to permutation data such as boundary orientation and symmetry class, while the polygon-diagram enumeration of
\cite{do2019countingnoncrossingpermutationssurfaces} extends naturally to arbitrary genus and number of boundaries. An interesting direction for future work is to develop a unified formalism that retains both features.

\acknowledgments{We thank Xuchen Cao, Tom Faulkner, and Kassahun Betre for fruitful discussions, and Henry Maxfield, Tom Hartman, and Kenneth Higginbotham for insightful conversations at an early stage of this project. This research is supported by the U.S. Department of Energy, Office of Science RENEW-HEP program under Award Number DE-SC0024518. N.P. was also supported by the National Science Foundation’s Research Traineeship (NRT) Fellowship funded by Grant No. DGE-2125906.}

\appendix
\section{Resolvents and the \texorpdfstring{$R$}{R}-Transform}

\label{app:resolvent-rtransform}

The main text compared noncrossing and annular noncrossing diagrams directly with the disk and double trumpet geometries of JT gravity, and argued that the diagrammatic description provides a complementary representation of the same large-$N$ matrix-model physics. In this appendix we sharpen this connection by reviewing mathematical transforms from one to the other. Random matrix correlation functions are usually expressed in terms of resolvents and spectral densities, while the diagrammatic description used here is naturally formulated in terms of moments and \textit{free cumulants}. Specifically, in free probability theory, the moment--cumulant relations and functional inversion relate the $R$-transform exactly to the resolvent.\footnote{This was a major triumph of the theory, showing that spectral and non-crossing permutation methods can be used to analyze random matrix models in the large-$N$ limit \cite{voiculescu_limit_1991,speicher_combinatorial_1998}. Free probability was originally developed in the study of operator algebras, in particular von Neumann algebras \cite{voiculescu_limit_1991,2014arXiv1404.3393S},
which have more recently played an important role in quantum gravity and holography
\cite{witten_why_2022,witten_gravity_2022,
chandrasekaran_large_2023,penington_algebras_2023,
kolchmeyer_von_2023,leutheusser_subregion-subalgebra_2024,
liu_lectures_2025}.} 

\subsection{Moments in Two Languages}

For an $N\times N$ random matrix $A_N$ with a large-$N$ limiting distribution, the normalized moments are
\begin{equation}
    m_n
    =
    \lim_{N\to\infty}
    \mathbb E\left[
        \frac{1}{N}\Tr A_N^n
    \right].
\end{equation}
These moments admit two equivalent descriptions. On the spectral side, if $\rho(x)$ is the limiting eigenvalue density, then
\begin{equation*}
    m_n
    =
    \int dx\ x^n\rho(x)\ ,
\end{equation*}
In free probability, moments are related to free cumulants through the non-crossing moment--cumulant formula
\begin{equation*}
    m_n
    =
    \sum_{\pi\in {\rm NC}(n)}
    \prod_{B\in\pi}
    \kappa_{|B|},
\end{equation*}
where the $\kappa_j$ are free cumulants and $B$ runs over the blocks of the partition $\pi$. Thus, the same moment can be described either spectrally, through an integral against an eigenvalue density, or combinatorially, through a weighted enumeration of non-crossing diagrams.

For JT gravity, these provide complementary descriptions of the same underlying data. The spectral description captures the density of energy states, while the combinatorial description captures the discrete structures whose continuum scaling produces the moduli-space quantities developed in this paper. Spectral information is encoded in the matrix model resolvent, whose double-scaling limit yields the JT spectral density. The corresponding moment and cumulant expansions are organized by non-crossing combinatorics, providing the discrete description underlying the
geometric basis for geometries in JT gravity.

\subsection{Resolvents}

The spectral properties of a matrix can be analyzed using the resolvent method, which at large-$N$ is given by

\begin{equation}
    G(z)
    =
    \lim_{N\to\infty}
    \mathbb E\left[
        \frac{1}{N}
        \Tr\frac{1}{z-A_N}
    \right],
\end{equation}
where $z$ is a complex number. Expanding at large $z$ gives
\begin{equation*}
    G(z)
    =
    \sum_{n=0}^{\infty}
    \frac{m_n}{z^{n+1}}\ ,
\end{equation*}
so the resolvent is the generating function for the moments of the distribution. It also determines the spectral density through its discontinuity across the cut \cite{wang_beyond_2023},
\begin{equation}
    \rho(x)
    =
    -\frac{1}{\pi}
    \lim_{\epsilon\to0^+}
    \operatorname{Im}G(x+i\epsilon)\ .
\end{equation}
Hence, the spectral edge is encoded in the analytic properties of $G(z)$, where a square-root edge of the density corresponds to a square-root branch point. At $\lambda=1$, for example, the Marchenko--Pastur law has density
\begin{equation*}
    \rho_{\rm MP}(x)
    =
    \frac{1}{2\pi}
    \sqrt{\frac{4-x}{x}},
    \qquad
    x\in[0,4]\ ,
\end{equation*}
and resolvent
\begin{equation}
    G_{\rm MP}(z)
    =
    \frac{z-\sqrt{z(z-4)}}{2z}\ .
\end{equation}
This has a square root branch-point at $x_\ast=4$.

\subsection{The \texorpdfstring{$R$}{R}-transform}

Free probability computes moments from non-crossing diagrams. The free cumulants $\kappa_n$ weight these diagrams, and the sum over them defines the moments of all orders through the moment--cumulant relation defined above. The generating function of the free cumulants is the $R$-transform,
\begin{equation}
    R(w)
    =
    \sum_{k\geq 1}
    \kappa_k w^{k-1}\ .
\end{equation}
All free cumulants of the Marchenko--Pastur law are equal, with $\kappa_k=\lambda$ for $k\geq1$, so its $R$-transform is
\begin{equation}
    R(w)
    =
    \frac{\lambda}{1-w}\ .
\end{equation}
At $\lambda=1$ every free cumulant equals one. The moment--cumulant relation then assigns unit weight to every noncrossing partition, so the $n$th moment is the number of noncrossing partitions of $n$ points, which is the Catalan number.

In this formalism, the disk topology appears because the moment--cumulant relation is a sum over noncrossing partitions drawn on a disk, as made explicit in Section~\ref{sec:nc-defn}. The large-$n$ limit of these diagrams is the combinatorial counterpart of the square-root edge limit seen in the resolvent. This can be made sharper by writing down the transformation that takes one formalism into the other.

\subsection{Functional Inversion}

The resolvent and the $R$-transform are related by functional inversion. Let $K(w)$ be the functional inverse of $G(z)$, so that $G(K(w)) = w$.
The $R$-transform is then defined by
\begin{equation*}
    K(w)
    =
    \frac{1}{w}
    +
    R(w) \ ,
\end{equation*}
and, equivalently, the resolvent satisfies the self-consistency equation
\begin{equation}
    G(z)
    =
    \frac{1}{
        z-R(G(z))
    }\ .
    \label{eqn:selfconsistent}
\end{equation}
Thus the $R$-transform, encoding information about non-crossing permutations through the free cumulant relations, and the resolvent provide equivalent descriptions.

\bibliographystyle{apsrev4-1long}
\bibliography{biblio2}

\end{document}